# FIM: A Fatigued-Injured Muscle Model Based on the Sliding Filament Theory

Fatemeh Jalali[a], Mohammad Ali Nazari[a,c*], Arash Bahrami[a], Pascal Perrier[b], Yohan Payan[c]

[a]School of Mechanical Engineering, College of Engineering, University of Tehran, Tehran, Iran

[b]Univ. Grenoble Alpes, CNRS, Grenoble INP, GIPSA-lab, 38000 Grenoble, France

[c]Univ. Grenoble Alpes, CNRS, Grenoble INP, TIMC, 38000 Grenoble, France

[*]Corresponding author: Mohammad Ali Nazari

Office No. 617, Biomechanics Department , School of Mechanical Engineering,
College of Engineering, University of Tehran
North Kargar Ave, Al-Ahmad crossroad
1439955961 Tehran, IRAN
Tel: +98 (0) 21 8800 5677 -Fax: +98 (0) 21 8801 3029
Office: +98 (0) 21 6111 5237
Email: manazari@ut.ac.ir

# Abstract
Skeletal muscle modeling has a vital role in movement studies and the development of therapeutic approaches. In the current study, a Huxley-based model for skeletal muscle is proposed, which demonstrates the impact of impairments in muscle characteristics. This model focuses on three identified ions: $H^+$, inorganic phosphate $P_i$ and $Ca^{2+}$. Modifications are made to actin-myosin attachment and detachment rates to study the effects of $H^+$ and $P_i$. Additionally, an activation coefficient is included to represent the role of calcium ions interacting with troponin, highlighting the importance of $Ca^{2+}$. It is found that maximum isometric muscle force decreases by 9.5% due to a reduction in pH from 7.4 to 6.5 and by 47.5% in case of the combination of a reduction in pH and an increase of $P_i$ concentration up to 30 mM, respectively. Then the force decline caused by a fall in the active calcium ions is studied. When only 15% of the total calcium in the myofibrillar space is able to interact with troponin, up to 80% force drop is anticipated by the model. The proposed fatigued-injured muscle model is useful to study the effect of various shortening velocities and initial muscle-tendon lengths on muscle force; in addition, the benefits of the model go beyond predicting the force in different conditions as it can also predict muscle stiffness and power. The power and stiffness decrease by 40% and 6.5%, respectively, due to the pH reduction, and the simultaneous accumulation of $H^+$ and $P_i$ leads to a 50% and 18% drop in power and stiffness.

## 1. Introduction

The modeling of a skeletal muscle is a fascinating field of study and has significant implications in understanding movement and developing therapeutic approaches. Multiple models have been proposed, with each model designed to accomplish specific objectives. Among the fundamental models, the Hill type models [1], the phenomenological models that focus on gross mechanical behavior, provide a functional view of muscle behavior for musculoskeletal simulation studies. However, the limited empirical explanation of these models come at the cost of not being able to offer precise predictions [2]. Therefore, more advanced models based on Hill models have been proposed that take into account the effect of calcium ions on muscle contraction. The proposed model in [3], based on the Hill model, enables the modeling of store-operated calcium entry (SOCE) channels and focuses on two primary chemical reactions comprising calcium release from sarcoplasmic reticulum and its binding to contractile proteins. Similar to this proposed model, other models [4, 5] based on the Hill model have also considered the effect of calcium on muscle contraction, predicting force responses for fast and slow-twitch muscles in response to a range of stimulation patterns, and determining the activation pattern to generate maximum force. In addition to the all mentioned models, the Huxley microscopic muscle models [6-11] based on the crossbridge kinetics have also emerged. These models provide a comprehensive and accurate description of muscle behavior, making them the preferred choice over other muscle models. Experimental validation of these models has helped to explain various muscle behaviors and phenomena [12].

The behavior of healthy muscles has been studied with various muscle models. However, modeling a defective muscle that cannot produce sufficient energy and force due to injury, fatigue, or atrophy is more complex and requires detailed metabolic pathways. In muscle energy production, adenosine triphosphate (ATP) is the immediate source of energy [13], and various ions such as $Ca^{2+}, Na^+, K^+$, and $Mg^{2+}$ are involved [14]. With muscle damage, new ions such as $H^+$ and inorganic phosphate $P_i$ play a critical role in the muscular force-producing capacity [13-15]. As a result, muscle models that can factor in the impact of ions are suitable for predicting injuries resulting from disorders in ion concentrations [4, 5] or energy production mechanisms required by cells [16-18]. Along with the models discussed above, geometric and topological data analysis can be used to assess the organizational signatures of skeletal muscle tissues for the diagnosis of various diseases, such as muscular dystrophies and neurogenic atrophies [19].

Our aim is to develop a muscle model that can accurately account for muscle disorders. To reach this goal, we began by studying skeletal muscle fatigue extensively. Muscle fatigue can stem from several minutes of intense muscle activity and results in a reduction in muscle force and power [20]. At the cellular level, this corresponds to changes in the excitation-contraction coupling, ionic alterations, and perturbations in cell metabolism [21]. The inhibition of the $Ca^{2+}$ release channel in the sarcoplasmic reticulum is a primary cause of changes in the excitation-contraction coupling. Consequently, reductions in $Ca^{2+}$ transition occur, leading to decreased muscle force and power [21]. Intense and short-term activity yields lactic acid build-up, which increases $H^+$ ion concentration and reduces the muscle force capacity. Acidosis lowers the pH, reducing the force produced by crossbridges. The $H^+$ ions further affect the $Ca^{2+}$ attachment to troponin C, reducing the number of active crossbridges. Low pH also slows down the rate of myosin attachment [22, 23] by inhibiting the $Ca^{2+}$ release channel and altering $Ca^{2+}$ reuptake [21]. Moreover, compared to near-physiological temperatures (30-32 °C), low temperature (10 to 15 °C) decreases force more significantly when the pH is reduced from 7 to 6.2. It has been observed that the increase of the concentration of $H^+$ and the consequent decrease in the pH from 7 to 6.2 in rabbits, mice, and rats specimens, induce a decrease of the muscle force at low temperatures by 28 to 53%, depending on the type of muscle fiber. At near physiological temperatures, the reduction in force is between 4 and 18% [24-26].

Intracellular $P_i$ accumulation due to ATP breakdown and creatine phosphate decomposition is another phenomenon that occurs in muscular fatigue due to during short-term and intense activities [27], leading to decreased maximal muscle force and crossbridge inhibition. A rise in $P_i$ concentration reduces the rate of $Ca^{2+}$ uptake from the sarcoplasmic reticulum and slows down the relaxation process [21, 28]. It also quickens the rate of myosin detachment from actin and prolongs myosin's attachment to actin by preventing ATP attachment to the myosin head [29]. It can be concluded that $Ca^{2+}$, $H^+$, and $P_i$ ions are the three primary agents responsible for muscle fatigue. In animals such as rabbits and rats, simultaneous increase in $P_i$ ions (up to 30 mM) and decrease in pH (up to 6.2) lead to maximum isometric force reduction by 36% in rat's slow fibers and 81% in rabbit's fast fibers [30-32]. These alterations in muscle behavior should impact muscle models' development. By integrating the above findings, it is possible to create more sophisticated and accurate models for a wide range of muscle conditions from healthy to injured ones.

Several models [3-5] have been proposed focusing on modeling the effects of calcium ions in the Hill model and the impact of these ions in the Huxley model [9, 11]. Additionally, the branched pathway model [16, 17] is capable of representing the effects of inorganic phosphate and hydrogen ions. However, reaching a more complete model that accounts for the simultaneous impact of all three ions presented two main challenges. The first challenge was selecting a model that can be modified to fit our desired fatigue model. The Hill model, being a phenomenological model, cannot account for the effects of $H^+$ and $P_i$ ions since their primary impact is on the rates of binding and dissociation of the contractile proteins, which this model does not take into account. Despite its usefulness in several applications, the Huxley model does not reflect the effects of these ions either. However, since the mechanisms of binding and dissociation of actin and myosin contractile proteins are considered in the Huxley model, this model appears to be a better candidate for modification according to our desired model. The second challenge is the computational costs associated with designing such a complex and comprehensive model. Fortunately, the method of Distribution Moment [7] has been developed that enables us to solve the Huxley equations with relatively simple calculations, and to obtain the information required for our model quickly and efficiently.

In this study, a microscopic model based on the theory of sliding filaments is proposed, hereafter referred to as a fatigue-injured muscle model (FIM model). The Huxley model is the kernel of the FIM. Three modifications were made in the last modified version of the Huxley model to embody the effect of $H^+$, $P_i$, and $Ca^{2+}$. The summation of four rates is replaced the attachment rate, three of which are fundamental force- and displacement-dependent reaction rates in the muscle contraction cycle, to consider the effect of $H^+$ and $P_i$. Since $P_i$ mainly alters myosin unbinding from actin sites, adjustments were made also to the detachment rate, in order to take the entire efficacy of $P_i$ into account. The Distribution Moment (DM) method [7] was applied to solve the Huxley equation.

## 2. Methods

The FIM model is based on the sliding filament theory (Huxley model). To include the effect of changes in $Ca^{2+}$, $H^+$, and $P_i$ ions, this model is enhanced using the Branched pathway model [17]. In this section, the Huxley model, which forms the basis of the FIM model, is presented.

Huxley [6] proposed a mathematical model consistent with the microscopic structure of muscle reproducing the macroscopic properties of muscle. In this model, it is assumed that each cross-bridge interacts with only one active site at each instant. If $h$ is the maximum distance that a cross-bridge can move to make contact with the active site, then the displacement of the cross-bridge from its equilibrium position is shown by $x$ [33]. When myosin extends to the size of $x$ at instant $t$ and attaches to actin-site, this attachment produces a force. The probability of being in this state is displayed by a distribution function $n(x,t)$. The following kinetics holds for this distribution (eq. 1) [33]:

$$\frac{\partial n(x,t)}{\partial t} - v(t)\frac{\partial n(x,t)}{\partial x} = f(x) - [f(x) + g(x)]n(x,t) \tag{1}$$

where $f$ is the rate by which the myosin attaches to the actin site (forward or binding rate), $g$ is the rate of detachment (backward- or unbinding rate), and $v(t)$ is the myofilament shortening velocity. The reason for the popularity of this theory is that once the distribution function $n$ is found, the contractile element's stiffness, force, and energy, can be found by computing the first, second, and third moments of $n$. If the bond length is normalized by $h$, the moments of the bond distribution function are expressed as:

$$Q_i(t) = 1/h \int_{-\infty}^{+\infty} (x/h)^i n(x/h, t)dx \quad i \geq 0 \tag{2}$$

To solve the Huxley kinetics equation (1), instead of solving the partial differential equation numerically, the Distribution Moment (DM) method is used [7]. In this method, the distribution function $n$ is assumed as a Gaussian distribution, which is not an exact account of the reality, but provides good estimates of the low-degree moments of the variable $n(x,t)$, which represent the structural, mechanical, and energy properties of the muscle. Using this approximation, the partial differential equation becomes three coupled ordinary differential equations (ODEs).

$$\dot{Q}_i = \beta_i - \phi_i - iu(t)Q_{i-1} \quad i = 0,1,2 \ \& \ Q_{-1} = 0 \tag{3}$$

where $u(t)$ is the myofilament's shortening velocity normalized by $h$, $\beta_i, i = 0,1,2 -$ are the attachment rates moments, and $\phi_i, , i = 0,1,2 -$ are the functions of both attachment and detachment rates and bond distribution. $\dot{Q}_i$ represents the time derivative of $Q_i$ (see Appendix for more details).

The following modifications have been done on the system of coupled differential equations (eq. (3)), to improve the final estimation:

1) Depending on the level of overlap between actin and myosin, only a certain portion of myosin can interact with the actin site, therefore the ratio of participating cross-bridges was introduced via a parameter called *alpha* [34]. The magnitude of *alpha* is obtained from an experimental force-length diagram of muscle-tendon [35]. The incorporation of alpha into the DM model occurs via the rate equations similar to [34], rather than the moment equations used for the bond distribution. This approach ensures that the cross-bridge binding rates account for the accurate number of participating crossbridges before integration takes place.

2) Since tendon stiffness, like many biological soft tissues, increases with applied strain, a *variable compliance* was defined for the tendon [10].

3) Tendon compliance is not constant at low loads but it is limited. To consider this limitation, another equation was added to the three ordinary differential equations of the DM model, which relates the velocity of the muscle-tendon unit ($\dot{L}_{MT}$) to tendon compliance, $K(Q_1)$, and $u(t)$ [10].

The aforementioned alterations are consolidated within the system of ordinary equations denoted as equation (3), resulting in the derivation of a final set of four ordinary differential equations (ODEs), which can be expressed as follows:

$$u(t) = \frac{L_{MT} s_0}{L_{M\_T(opt)} h} (\dot{Q}_1 K(Q_1) - \frac{\dot{L}_{M\_T}}{L_{M\_T(opt)}}) \qquad (4)$$

$$\dot{Q}_i = \alpha \beta_i - \phi_i - i u(t) Q_{i-1} \quad i = 0,1,2 \ \& \ Q_{-1} = 0 \qquad (5)$$

In equation (4), $L_{M\_T}$ is the current length of the muscle-tendon unit, $s_0$ is sarcomere length at reference condition, and $L_{M\_T(opt)}$ represents the optimum length of the muscle-tendon complex.

All the modifications mentioned above were considered in the Huxley model, and after achieving a more accurate model, modifications were made step by step to add the effect of the three above mentioned essential ions in fatigue. To incorporate the effects of $H^+$, and $P_i$ ion, it was necessary to modify the actin-myosin binding and detachment rates. Finding a substitute for the linear relationship that represents the binding rate in the Huxley model [6] is of great importance. Among the microscopic models that consider the actin-myosin binding and detachment process step by step, the Walcott model [36], which is an earlier version of the branched pathways model [17], can effectively represent the mechanism of muscle contraction and the interaction between actin and myosin. Before discussing the Walcott and branched pathways models, it is essential to define how the binding and unbinding rates are defined in the Huxley model (section 2.1). In the following, we first describe the concept of binding and unbinding rates in the Huxley model and then explain how to substitute them with the binding and unbinding rates in the Walcott and branched pathways models in sections 2.3 and 2.4.

### 2.1. The attachment and detachment rates in the Huxley model

Figure 1 illustrates an overview of the interaction cycle between actin and myosin contractile proteins in the Huxley model, with darker lines showing the dominant reaction pathways. M, A, T, and D denote Myosin, Actin, ATP, and ADP (ADP+P), respectively. It is assumed that there is only one main unattached state (state (2) or (3) or their mixture). Similarly, all the states of Fig. 1. A where the myosin head is attached to the actin-site or the fast equilibrium mixture of successive states including at least one bounded actin and myosin (state (6), (4) or (5) or a mixture of them) is assumed as the only significant attached state [37]. Fig. 1. B shows the final rates. $f$ is the attachment rate, and $f'$ is its inverse, $g$ is the rate of detachment, and $g'$ is its inverse [9]. It has been found that $g'$ is very small and negligible

The interaction rates between myosin and actin can be divided into three distinct regions: the primary region, where the springtail of the myosin is compressed and the myosin is not attached to the actin site ($-\infty < x/h < 0$); the secondary region, where the myosin is attached to the actin site and extended by an amount of x ($0 < x/h < 1$); and the tertiary region, where the myosin springtail has been extended and detached from the actin site ($1 < x/h < \infty$). Each of these regions is characterized by its own attachment and detachment rates, denoted as $g_i$s and $f_i$s in equations (6) and (7), respectively.

$$g(x/h) + f'(x/h) = \begin{cases} g_2 & -\infty < x/h < 0 \\ g_1 x/h & 0 < x/h < 1 \\ g_1 x/h + g_3(x/h - 1) & 1 < x/h < \infty \end{cases} \quad (6)$$

$$f(x/h) = \begin{cases} 0 & -\infty < x/h < 0 \\ f_1 x/h & 0 < x/h < 1 \\ 0 & 1 < x/h < \infty \end{cases} \quad (7)$$

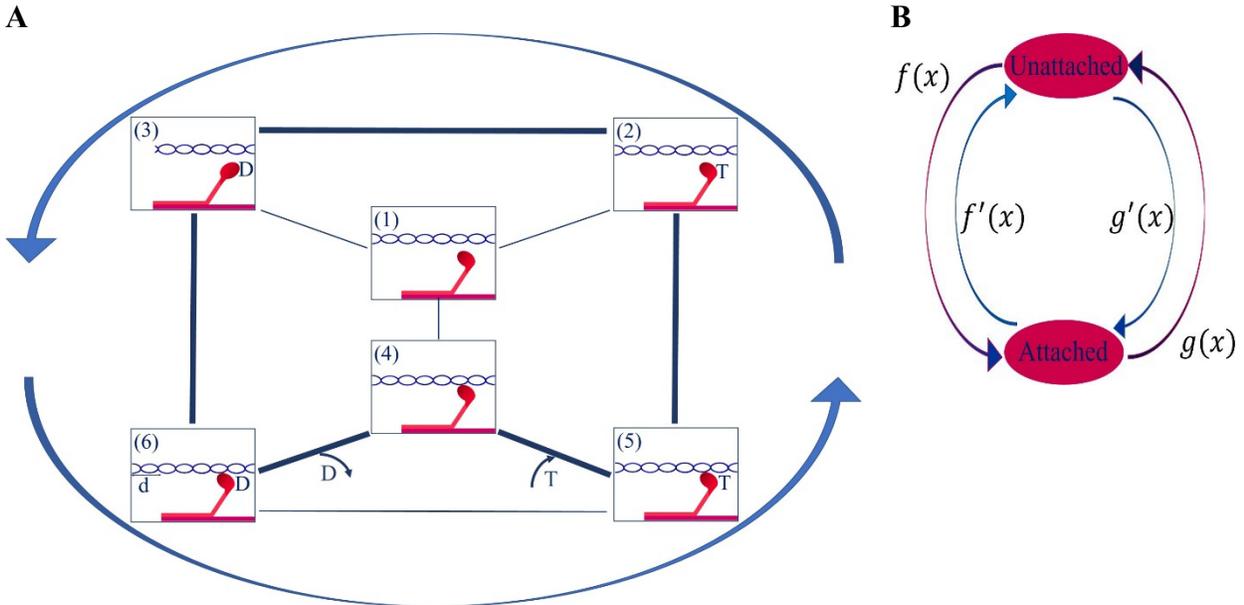

**Fig. 1. A**: The diagram shows different biochemical states of a cross-bridge, with myosin heads depicted as ovals and actin as chains. The letters T and D represent ATP and ADP (+P), respectively. The actin's displacement when it is pulled through myosin attachment is equal to *d*. Each line is representing two inverse feasible first-order transitions. Heavy lines show the dominant cycle, which involves five states. The general direction of the reaction is counterclockwise. **B**: The reduced two-state model: in this model, only one primary unattached state, either state (2) or state (3), or a mixture of them, exists. Furthermore, all physiological states of **Fig. 1. A**, where the myosin head is bound to the actin-site or a mixture of successive states including at least one bounded actin and myosin, namely state (6), (4) or (5), or a mixture of them, are regarded as the only significant attached state. *f* is considered as the significant attachment rate, and its inverse is $f'$, while *g* represents the primary rate of detachment and $g'$ is its inverse [37]. State (1) occurs very fast hence it is not considered in this model.

## 2.2. Attachment rates in the Walcott model

In Fig. 2, both the Walcott [36] and branched pathway [17] models are presented, with the blue frames highlighting the distinctions between them. Initially, we will elucidate the stages of muscle contraction in the Walcott model, which is a microscopic-based model utilized for healthy muscles. The muscle contraction steps in the Walcott model [36] are as follows (Fig. 2). In the first step, the ATP (denoted as T in Fig. 2) hydrolysis products, ADP (denoted as D in Fig. 2) and inorganic phosphate ions ($P_i$) (denoted as P in Fig. 2), are on the active site of myosin (state (1) in Fig. 2), then $P_i$ is released and myosin is attached to the actin with the rate of $f_a(x)$ which is a function of displacement $x$. In the second step, the actin's displacement is equal to $d$ (state (2)) and in the following step, in a force-dependent process, the ADP is released from the myosin (state (3)) at a rate of $f_D(F)$. In the final step, once the ATP is attached to the myosin, the bond between actin and the myosin is detached (state (4)). The rate of binding of myosin to actin, i.e., the transition from state (1) to state (2), is estimated by following the Gaussian density function. The symbol $k$ represents the stiffness of myosin, which denotes the rigidity of myosin's lever arm in the context of being assumed as a linear spring This constant can be estimated through a laser trap assay [16].

$$f_a(x) = f_a \sqrt{\frac{kh^2}{2\pi K_B T}} \exp\left(-\frac{kx^2}{2K_B T}\right) \qquad (8)$$

The transition from state (2) to state (3) is estimated using Bell's equation [38]:

$$f_D(F) = f_D^0 \exp\left(-\frac{F\varepsilon_x}{K_B T}\right) \qquad (9)$$

where $f_D^0$ is the ADP release rate when the force is zero, $\varepsilon_x$ is a distance parameter that determines the force dependence. This constant can be found through a laser trap assay [16]. $K_B$ is the Boltzmann constant, and $T$ is the absolute temperature.

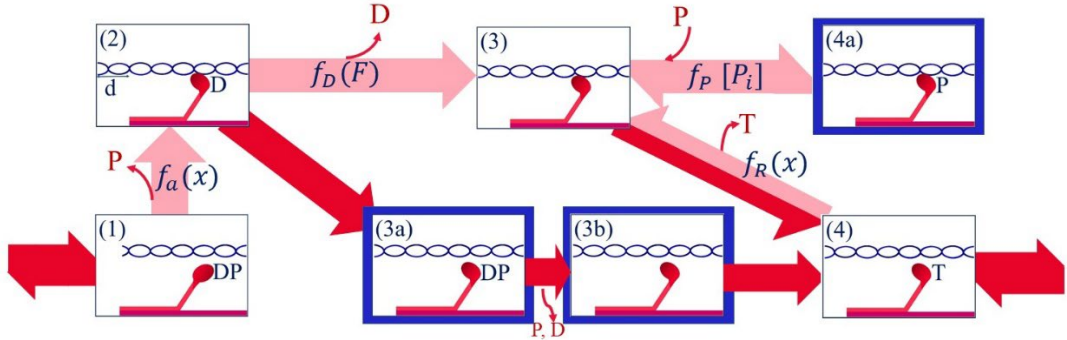

**Fig. 2.** The branched pathway model [17] with all the rates in which myosin is going to attach to actin or already has been attached to actin (significant attached state) is a more complete example of the Walcott model [36]. It demonstrates the impact of inorganic phosphate ions $P_i$ in the contraction process as a result of muscle fatigue, the additional stages are displayed in blue frames. In the Walcott model [36], the hydrolysis products of ATP- ADP and $P_i$ ions (marked with DP) are present on the active site of myosin (state (1)). Next, the $P_i$ ion is released, and myosin attaches to actin at a rate $f_a(x)$, which is a displacement (x) dependent process. At state (2), actin's displacement is equal to '$d$'. Through a force-dependent process, the ADP (marked with D) releases from the myosin at state (3), with a rate of $f_D(F)$. Finally, once the ATP (marked with T) binds to the myosin, the bond between actin and myosin disconnects, and the process enters the state (4). The impact of inorganic phosphate (Pi) on the cross-bridge cycle remains a point of contention. While it is widely agreed that $P_i$ diminishes muscular force by rebinding to myosin molecules that are strongly attached to actin, there is some evidence that $P_i$ can also enhance actin filament velocity in an acidic environment. To address this variable impact of $P_i$, the branched pathway model includes two alternative states for contractile proteins in the contraction process period (transitioning from state 2 to state 3, or to 3a and then to 3b). This means that when $P_i$ ions bind to myosin, it causes myosin to immediately detaches from its position after completing the powerstroke, without the reversal of the powerstroke. Another consequence of higher levels of $P_i$ concentration is that it can compete with ATP and prevents myosin detachment, resulting in myosin staying bound to actin for a longer period (transition between states 3 and 4a). $f_P[Pi]$ is the summation of $P_i$ binding and unbinding rates to myosin in the rigor state (state (3)).

It has been found that an increase in the concentration of $P_i$ ions affects the attachment and detachment rates of actin and myosin. These changes in rates are accounted for in the branched pathway model. They have been obtained in normal and acidic conditions as well as in the absence of inorganic phosphate ions and in the presence of these ions due to fatigue [16, 17] (Table 1). The branched pathway [17] model is based on the Walcott model, in which steps 3a, 3b, and 4a have been added to consider the effect of the $P_i$ concentration changes in fatigue condition on attachment and detachment rates (Fig. 2).

**Table 1.** Parameter values used in the FIM Model (from [16, 17][1])

| Parameter | Values in healthy condition | Values in acidosis |
| --- | --- | --- |
| pH | 7.4 | 6.5 |
| $f_a$ | $40 \ s^{-1}$ | $10 \ s^{-1}$ |
| $f_D^0$ | $350 \ s^{-1}$ | $100 \ s^{-1}$ |
| $f_R$ | $0 \ s^{-1}$ | $10 \ s^{-1}$ |
| $g_{off}$ | $30 \ mM^{-1} \ s^{-1}$ | $100 \ mM^{-1} \ s^{-1}$ |
| $f_P^+$ | $0.7 \ mM^{-1} \ s^{-1}$ | $0.7 \ mM^{-1} \ s^{-1}$ |
| $f_P^-$ | $10 \ s^{-1}$ | $10 \ s^{-1}$ |
| $g_{DP}$ | $40 \ s^{-1}$ | $40 \ s^{-1}$ |

[1]The last four parameters are added to the model only in the presence of $P_i$ ions. $g_{off}$ is the rate at which myosin detaches from actin in the presence of ADP due to the accumulation of $P_i$, $g_{DP}$ is the rate at which ADP and $P_i$ are detaching from the active site of myosin, $f_P^+$ and $f_P^-$ are the rate of $P_i$ attachment to myosin in the rigor state and its reverse rate respectively (see Appendix for more details).

## 2.3. Modification of the forward binding rate

The binding rate in the Huxley model was assumed to be the sum of the rates of occurrence of all states in which the cross-bridge was connected to or was going to be connected to the actin site. This rate is replaced by four rates in our FIM model:

1. the rate of myosin transition from detached to strongly bound state ($f_a(x)$, transition from state (1) to (2)),
2. the rate of ADP release by myosin when it is strongly bound to actin ($f_D(F)$, transition from state (2) to (3)),
3. myosin re-binding rate to actin ($f_R(x)$, transition from state (4) to state (3))
4. the summation of inorganic phosphate binding and unbinding rates to myosin active-site in the rigor state ($f_P[P_i]$, that belongs to the transition from the state (3) to 4a, and its reverse reaction where myosin has already been attached to actin).

The $f_R(x)$ and $f_a(x)$ rates areassumed to be Gaussian density functions [17].The reason for this choice is that the Gaussian distribution function assumption is common due to the central limit theorem [36]. The net binding rate function, $f$, is expressed as follows:

$$f(x, F, [P_i]) = f_a(x) + f_D(F) + f_R(x) + f_P[P_i] \qquad (10)$$

or:

$$f(x, F, [P_i]) = f_a \sqrt{\frac{kh^2}{2\pi K_B T}} \exp\left(-\frac{\kappa x^2}{2 K_B T}\right) + f_D^0 \exp\left(-\frac{F\varepsilon_x}{K_B T}\right)$$
$$+ f_R \sqrt{\frac{\kappa h^2}{2\pi K_B T}} \exp\left(-\frac{\kappa x^2}{2 K_B T}\right) + f_P[P_i] \qquad (11)$$

To use these functions in the DM model, they have been normalized with $h$ (the maximum distance a cross-bridge can move to attach to the actin-site). Then, they have been fitted in domain $0<x/h\leq1$ with proper polynomials ($R^2 \geq 0.994$) (see Appendix for more information).

## 2.4. Modification of the detachment rate

As a consequence of the phenomena that are accounted for in the branched model, the detachment rate in the Huxley model has to be modified. It should be noted that the rate of $g_1$ corresponds to the detachment rate for $\xi$ between zero and one. Since myosin binding to actin also occurs in this range, its modified value ($g_1 + 0.5g_{off}$) should not be much greater than the summation of the attachment rates, i.e., $f$. Relation (4) for the detachment rate is replaced with the following equation:

$$g'(x/h) = \begin{cases} g_2 & -\infty < x/h < 0 \\ (g_1 + 0.5g_{off})x/h & 0 < x/h < 1 \\ g_1 x/h + (g_3 + (c_{P_i} - 0.5)g_{off} + g_{DP})(x/h - 1) & 1 < x/h < \infty \end{cases} \qquad (12)$$

where $g_{off}$ is the rate at which myosin separates from actin taking into account the effect of $P_i$ binding when myosin is tightly bound to actin, $g_{DP}$ is the rate of release of inorganic phosphate and ADP from the active site of myosin when myosin is not bound to actin, and $c_{P_i}$ is the concentration of inorganic phosphate accumulated in muscle.

2.5. Consideration of activation mechanism induced by calcium ions

Due to the important role of calcium in muscle activation, different muscle models have been proposed to account for it [4, 5, 9, 11]. In their 1991 paper, Ma et al. [9, 11] coupled the model of muscle contraction with the dynamics of calcium activation by simply adding another assessable variable, namely the free calcium concentration. This enabled them to measure muscle force and energy due to variations in calcium concentration.

Initially, there are two proposed hypotheses for the kinetics of the myosin-actin-troponin complex [9]: tight and loose coupling. The key difference between them is that in the loose coupling scheme, calcium can bind and unbind from troponin regardless of the bonding state between actin and myosin, while in the tight coupling scheme, a troponin molecule can only release its bound calcium ions when the associated cross-bridge is detached from the actin. Given the compatibility of the tight-coupling hypothesis with experimental evidence [39], we decided to use this hypothesis in the FIM model.

The effect of calcium activity on muscle contraction is accounted for by introducing an activation coefficient, $r$, to the DM equations (eq. (3)). The variable $\phi$, as represented in equation (3), is currently divided into two distinct components based on the rate of binding in the forward and backward directions, respectively. $\phi_{1i}$ is a function of forward binding rate and bond distribution in the $\phi_i$ and $\phi_{2i}$ is dependent on the backward binding rate. (See Appendix for more details.)

$$\dot{Q}_i = r(c)\beta_i - r(c)\phi_{1i} - \phi_{2i} - iu(t)Q_{i-1} \quad i = 0,1,2 \ \& \ Q_{-1} = 0 \quad (13)$$

$$r(c) = \frac{c^2}{c^2 + \mu c + \mu^2} \quad (14)$$

where $c$ is cytosol calcium concentration and $\mu$ is the calcium-troponin equilibrium constant. Another equation was added to the equations for muscle contraction (eq. 13), total calcium concentration ($C$) relation (eq. 15, b is a structural parameter for crossbridges):

$$C(c, Q_0) = c + 2bQ_0 + r(c)\left(2 + \frac{\mu}{c}\right)(1 - bQ_0) \quad (15)$$

The activation coefficient in the FIM model is applied differently from the original model [11]. The purpose of our model is to investigate disturbances in free calcium concentration. For a specific amount of the ratio of free calcium concentration to total calcium ($\frac{c}{C} = ca\_ratio$), and by replacing the relation of the activation coefficient from eq. (14) in (15), the free calcium concentration is calculated (equations (16), and (17)). Having computed the concentration of free calcium, the activation coefficient is obtained. Using the activation coefficient in the model, muscle stiffness, force, and elastic energy can be calculated.

$$\frac{c}{ca\_ratio} = c + 2bQ_0 + \frac{c^2}{c^2 + \mu c + \mu^2}\left(2 + \frac{\mu}{c}\right)(1 - bQ_0) \qquad (16)$$

$$0 = c \times ca\_ratio - c + 2bQ_0 \times ca\_ratio + \frac{ca\_ratio \times c}{c^2 + \mu c + \mu^2}(2c + \mu)(1 - bQ_0) \qquad (17)$$

### 2.6. Modeling force generation in injured muscles

Muscle injuries such as laceration, contusion, and strain-induced damage are caused by mechanical trauma and have a permanent effect on muscle force and power [40, 41]. The difficulty in measuring the effect of injuries on the mechanical capacities of the muscle is one of the reasons why there is almost no model that can predict muscle force after partial- or complete-laceration.

After a laceration injury, the extent of muscle innervation damage and the degree of membrane integrity are two important parameters that determine the rate of recovery [40, 42-44]. Muscle regeneration after laceration is a complex and relatively long process that takes 56 days or even 12 weeks [40, 41, 45]. The tension in a fully transected muscle is 46% lower, and in a partially transected muscle, it is 38% lower. The ability of the muscle to shorten is much less affected. In a muscle that has been partially transected and re-sutured, the stitched part becomes atrophic if it is not innervated, and its ability to produce force will be reduced [40].

This effect is modeled in the FIM model via the activation coefficient (see Appendix for details) in the FIM model. Studying the consequences of variations of this activation coefficient enables us to investigate muscle force generation capacity in injured muscles, associated with changes in the ratio of free cytosol calcium to total calcium.

### 2.7. Numerical simulations

Numerical simulations of the proposed model were performed for healthy and fatigued muscles. The isometric force of muscle in fatigue conditions at the physiological temperature has been studied in several articles [24-26, 31]. The isometric force is first simulated in two states of fatigue due to pH reduction and accumulation of $P_i$ ions. Temporal variation of force and its slope are then obtained. To evaluate the importance of different attachment rates, a sensitivity analysis is performed. In all numerical simulations, it was assumed that the percentage of free calcium that is able to bind to troponin is 74.5% (the approximate ratio of $Ca^{2+}$ attached to Troponin C, the calcium-binding subunit of the troponin, to $Ca^{2+}$ release) [46] and the activation coefficient is 0.98. After model validation for healthy and fatigued muscle, the effect of calcium on isometric muscle force is discussed to predict the force of the injured muscle. For isometric force modeling, first, the muscle is fully activated under an instantaneous stimulation and then the muscle force is examined under conditions where this activation is constant. Detailed mathematical descriptions of our model are included in the Appendix.

The model equations were solved by the 4$^{th}$ and 5$^{th}$ order Runge-Kutta method in MATLAB software by using the ode45 function.

## 3. Results

In the Huxley model [6] the rate of myosin binding to the actin site , $f$, has not been determined experimentally and was hypothetically estimated to be able to predict mechanical and energy experimental results [4]. Using the procedure described in the method section, the forward binding rate in our FIM model was replaced by the summation of four rates. Three rates are related to the main myosin-to-actin bond reactions that were a function of myosin displacement and force. And the last one is the summation of inorganic phosphate binding and unbinding rate to myosin active-site in the rigor state. All parameters related to these rates can be calculated by fitting available experimental data [16, 17]. Therefore, the integrals of the $f$-rate in the bond interval, that is, for $x/h$ varying between zero and one in the Huxley and FIM models, must be comparable. After calculating the integral, the obtained value is 45.19. This value has very good compatibility with the values used in models based on the Huxley model, such as the Zahalak's model [7], i.e., 43.3, or the Force Depression Corr model [34], i.e., 50. In addition, the time-force diagram of the model is consistent with the time-force diagram of the Huxley model. The difference in initial slope between the two models is negligible, and eventually, both models converge to the force at optimum length (Fig. 3).

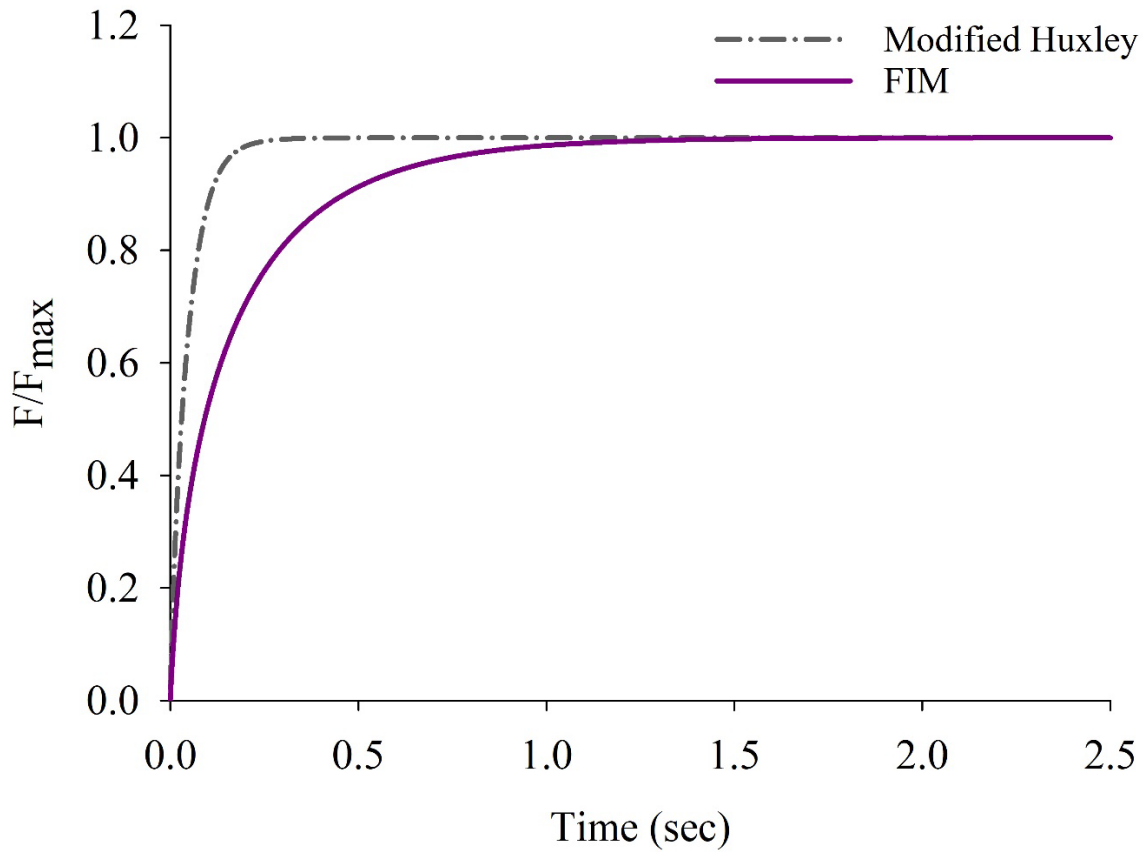

**Fig. 3.** Normalized force for modified Huxley (dot-line) and FIM (solid line) models vs. time. The values of parameters for the modified Huxley model have been extracted from Corr et al. [34], the values of parameters for FIM are the same except for actin-myosin binding rates which are [17]: the rate of myosin transition from detached to strongly bound state $(f_a)$ = 40 $s^{-1}$, ADP release rate $(f_D^0)$ = 350 $s^{-1}$, and myosin to actin rebinding rate $(f_R)$ = 0 $s^{-1}$. The force in each model is normalized by the muscle force at the optimum muscle-tendon length. The difference in their behavior comes from binding rate functions, which in FIM are exponential with respect to the linear function in modified Huxley.

3.1.          Maximum isometric force

In a study of New Zealand white rabbit psoas muscle, it had been found that lowering the pH from 7 in the control muscle to 6.2 in the acidic condition at 30 °C reduced the isometric tension by 17.8% [24]. In another study of the flexor brevis muscle in male mice, it was recognized that lowering the pH at physiological temperature (32 °C) reduced the isometric force of the muscle by about 10% [25]. A more recent study of rat soleus and gastrocnemius muscle also showed that under acidic conditions (pH 6.2), fiber force was reduced by 4-12% [26]. Replacing the parameters of acidosis in the FIM model decreases the isometric force. It can be seen that the model shows a 9.5% reduction in isometric muscle force under acidic conditions at physiological temperatures. This result is very well compatible with the experimental data.

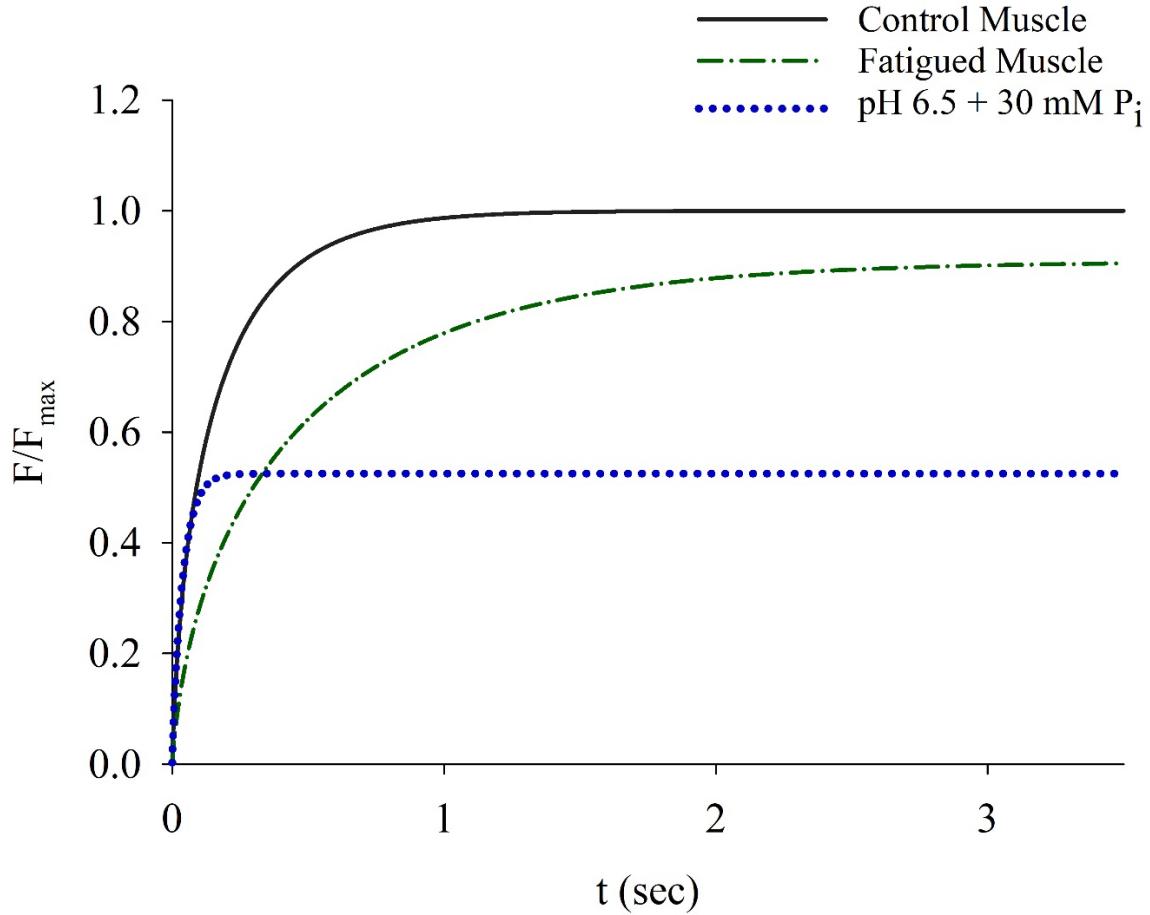

**Fig. 4.** The force-time curve for a control muscle (solid line), a fatigued muscle due to a drop in pH from 7.4 to 6.5 (dash-dotted line), and fatigue resulting from the combined effect of a pH drop (from 7.4 to 6.5) and an accumulation of 30 mM of inorganic phosphate $P_i$ (dotted line). The values of the myosin transition rate from detached to strongly bound state ($f_a$), ADP release rate ($f_D^0$), and myosin rebinding rate to actin ($f_R$) for a healthy muscle are $40\ s^{-1}$, $350\ s^{-1}$, and $0\ s^{-1}$, respectively [17]. The detachment rates are reported in [34]. For the fatigued muscle due to a drop in pH, the rates of $10\ s^{-1}$, $100\ s^{-1}$, and $10\ s^{-1}$ for myosin transition from detached to strongly bound state, ADP release, and myosin rebinding to actin, respectively, are used [17]. The values of the detachment rate are similar to those for a healthy muscle (control muscle) [34]. Due to the pH drop the normalized isometric force reduces by 9.5%. The rates for a muscle subjected to a pH drop from 7.4 to 6.5 and simultaneous inorganic phosphate accumulation (30 mM) are as follows [17]: $f_a = 10\ s^{-1}$, $f_D^0 = 100\ s^{-1}$, $f_R = 10\ s^{-1}$, $g_{off} = 100\ mM^{-1}\ s^{-1}$, $f_P^+ = 0.7\ mM^{-1}\ s^{-1}$, $f_P^- = 10\ s^{-1}$, and $g_{DP} = 40\ s^{-1}$. The rate at which myosin detaches from actin due to the binding of $[P_i]$ when myosin is strongly bound to actin is referred to as $g_{off}$. The rate at which $P_i$ and ADP are released from the active site of myosin when it is not bound to actin is known as $g_{DP}$. Furthermore, $f_P^+$ denotes the reaction rate for the release of inorganic phosphate from myosin while it is bound to actin, and $f_P^-$ represents the reverse rate of this reaction (see Appendix for more details). The simultaneous accumulation of $H^+$, and $P_i$ ions causes a 47.5% reduction in normalized isometric force.

The time constant has increased by 72.7%. One of the advantages of numerical modeling, assuming that it is correct, over experimental analysis is the possibility of predicting a behavior over a wider range of data. Figure 5 shows the force-time diagrams of a muscle that is shortened with different magnitudes for both the control muscle and the fatigued muscle in isometric conditions.

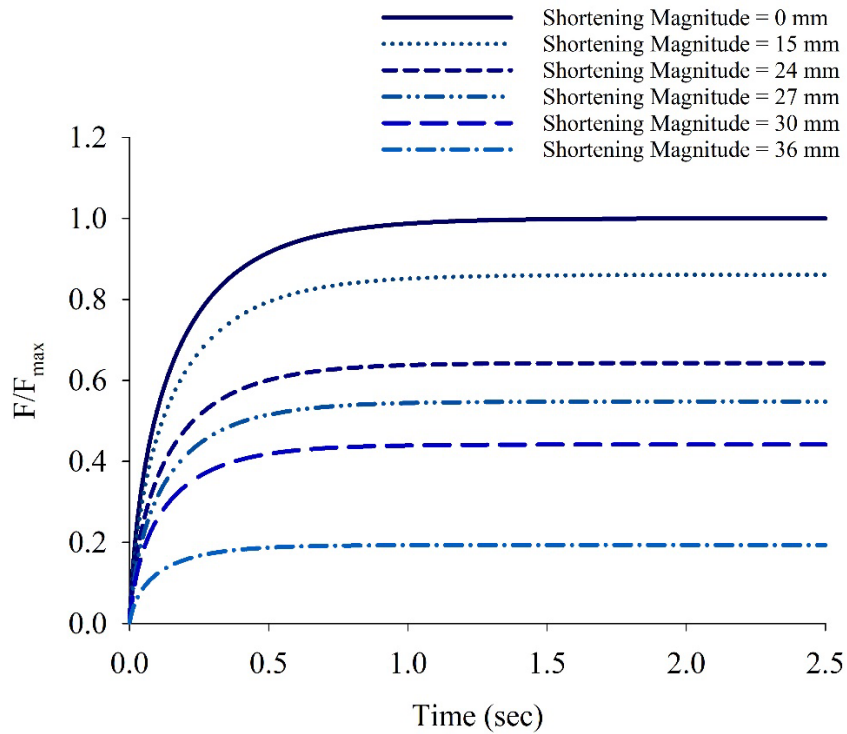

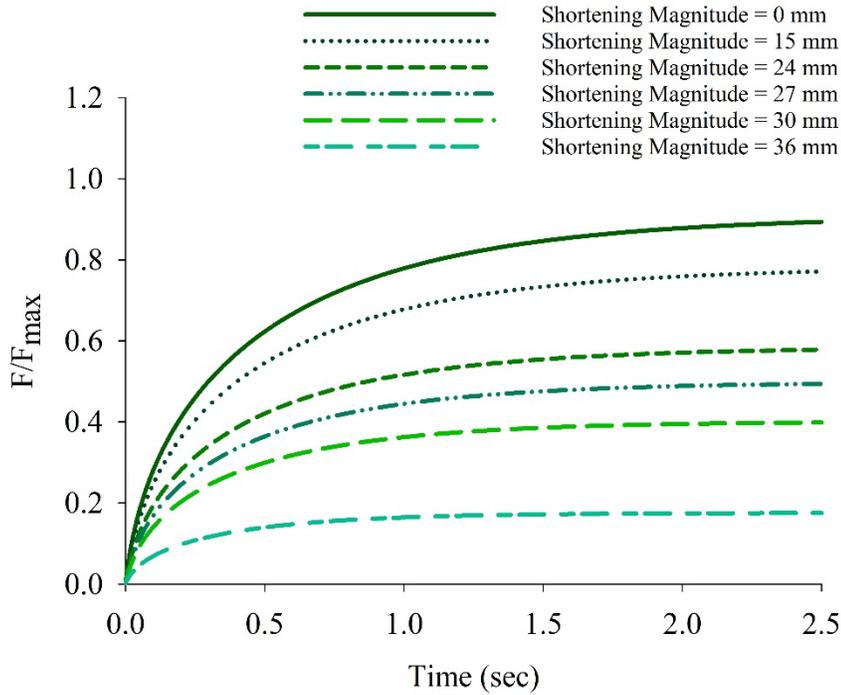

**Fig. 5.** Model-based predicted force-time responses of a muscle that has been subjected to different shortening magnitudes for both control muscles (top) and fatigued muscles (bottom). When a muscle is under maximum isometric contraction at different lengths, the fatigued muscle shows a large reduction in isometric force compared to control muscle. In all diagrams, the tendon-muscle velocity is zero, and the muscle produces force due to stimulation caused by its initial frequency. The values of myosin transition rate from detached to strongly bound state ($f_a$), ADP release rate ($f_D^0$), and myosin rebinding rate to actin ($f_R$) for a healthy muscle are $40\ s^{-1}$, $350\ s^{-1}$, and $0\ s^{-1}$, respectively [17]. The detachment rate values are according to Corr et al. [34]. For the fatigued muscle due to a drop in pH, the corresponding rates of $10\ s^{-1}$, $100\ s^{-1}$, and $10\ s^{-1}$ for the three mentioned rates ($f_a$, $f_D^0$, and $f_R$), are used [17]. The detachment rates are similar to those for healthy muscles (control muscle) [34].

As the shortening magnitude increases, the initial slope decreases, and the time constant increases. In a certain shortening magnitude, the force produced in the muscle in the fatigue condition is less than the force in the control muscle.

One of the remarkable capabilities of our FIM model is the prediction of force in voluntary motions with different shortening velocities. The length-force diagram is plotted at zero velocity and two voluntary speeds of 0.5 and 1.5 mm/s. As can be seen in Fig. 6, as the shortening velocity increases, the force decreases, and this force reduction is larger in the fatigue condition than in the control state.

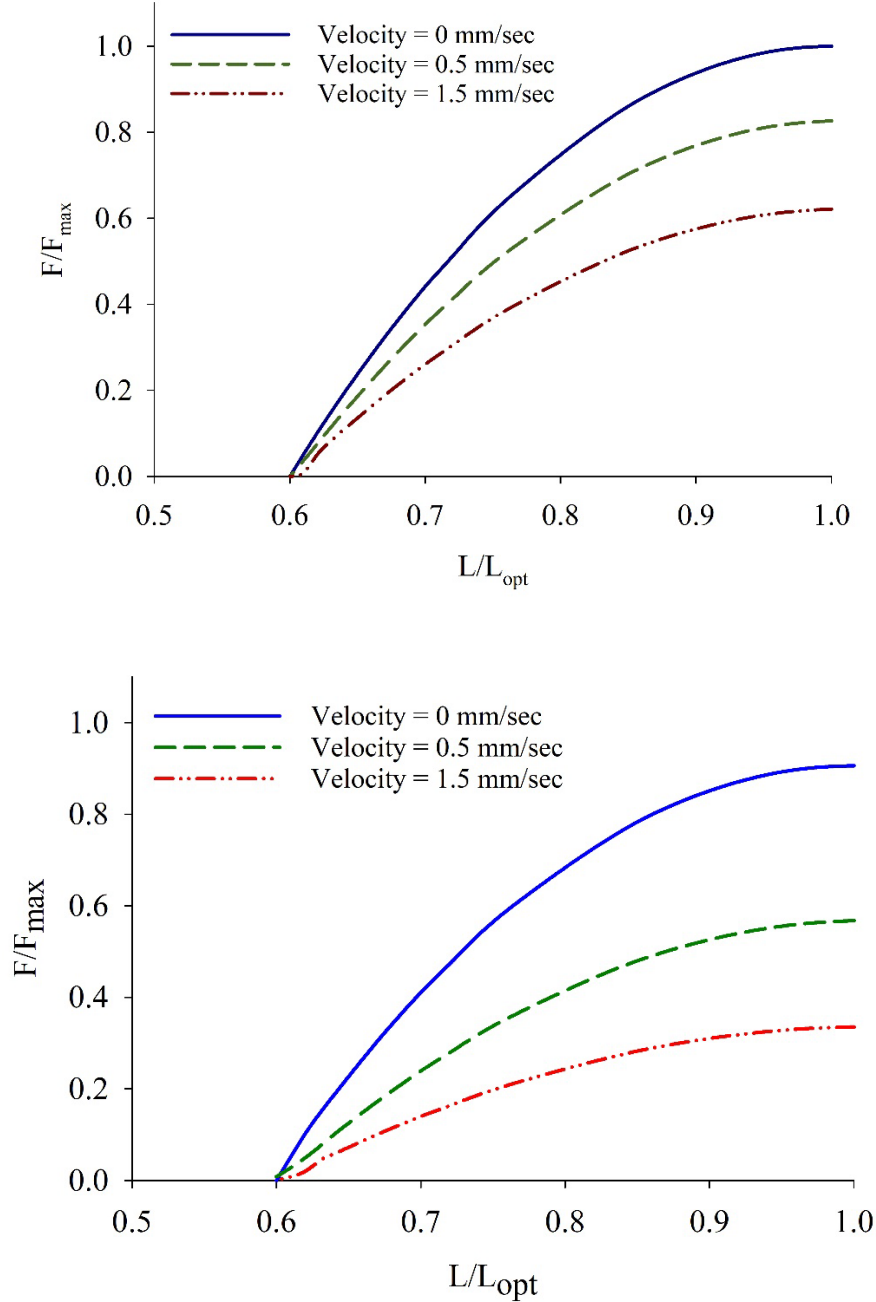

**Fig. 6.** Normalized force vs. length diagram in control muscle (top) and fatigued muscle due to pH drop from 7.4 to 6.5 (bottom). As the rate of shortening changes in the muscle, it is observed that the higher the shortening velocity, the less force is produced by the muscle in both control and fatigue states. the rates of a healthy muscle's myosin transition from detached to strongly bound state ($f_a$), release of ADP ($f_D^0$), and rebinding of myosin to actin ($f_R$), are $40\ s^{-1}$, $350\ s^{-1}$, and $0\ s^{-1}$, respectively. The detachment rate values are provided from [34]. When the muscle is fatigued due to a decrease in pH, these rates become $10\ s^{-1}$, $100\ s^{-1}$, and $10\ s^{-1}$ for $f_a$, $f_D^0$, and $f_R$, respectively, while the detachment rates remain similar to those for healthy muscles, as reported in [34].

### 3.2. The simultaneous effect of acidosis and accumulation of $P_i$ ions

In a study of rabbit's psoas muscle, Karatzaferi et al. [30] found that the simultaneous increase in $P_i$ to 30 mM and decrease in pH to 6.2 at 30 °C reduced isometric tension by 52% ($from\ 215 \pm 10\ to\ 103 \pm 9\ \frac{kN}{mm^2}$). Nelson et al. [32] also examined the type I, IIa, and IIx fibers of male and female rats in conditions similar to Karatzaferi et al. in terms of temperature, pH, and $P_i$ concentration. They observed that the peak force was reduced by 44% in type I fiber (the soleus) and by 41% and 50% in type IIa (the deep region of the lateral head of the gastrocnemius) and IIx fibers (the superficial region of the medial head of the gastrocnemius), respectively. In another study by Nelson et al. [31] on type I and IIx fibers of male rats, it was found that the simultaneous decrease in pH to 6.2 and 30 mM increase in $P_i$ could reduce the maximum isometric force of type I fiber by 36%, and reduce it by 46% in type IIx fibers. After modifying the detachment rates and applying the effect of pH reduction in the FIM model, it was observed that the force is reduced by 47.5% (Fig. 4). The accumulation of $P_i$, like the one of hydrogen ions, reduces the force such that the simultaneous accumulation of the two factors intensifies the reduction of force.

### 3.3. An analysis of the effect of the variations of three main binding rates ($f_a$, $f_D^0$, and $f_R$)

Considering the fatigue conditions for rates $f_a$, $f_D^0$, and $f_R$, the effect of each of these three rates was examined by keeping the other two rates constant. As can be seen from Fig. 7, the reduction in the release rate of ADP from myosin has a larger effect on the isometric force than the rate of rebinding to the rigor state and the rate of binding from the detached to the strongly-bond state. By changing the rates $f_D^0$ and $f_a$, the total attachment rate decreases, and conversely, by changing the rate $f_R$, the total binding rate increases, but this increase is very small compared to the decreasing effect of the other two rates.

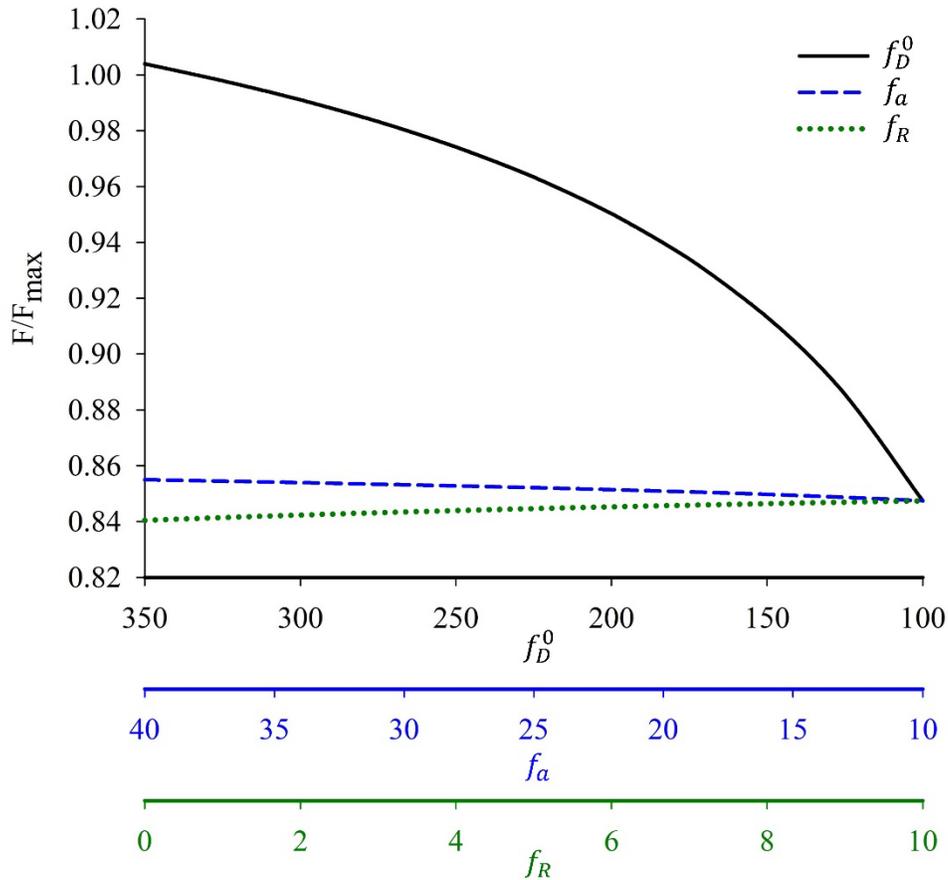

**Fig. 7.** Due to acidosis, $f_D^0$ (ADP release rate) decreases from 350 to 100 $s^{-1}$ and $f_a$ (the rate of myosin transition from detached to strongly bound state) from 40 to 10 $s^{-1}$. Instead, the $f_R$ (myosin rebinding rate to actin) increases from 0 to 10 $s^{-1}$. To investigate the effect of each of these three rates, two rates have been kept constant at the value corresponding to the fatigue state, and the third rate has been changed in its range during acidosis. For example, in the first row, the $f_a$ and $f_R$ values are assumed to be 10 and 10 $s^{-1}$, respectively, and the $f_D^0$ value is reduced from 350 to 100 $s^{-1}$. Like other graphs, force is normalized by muscle force with respect to the optimum length of tendon-muscle, and $f$ is normalized by the amount of binding rate at normal conditions (pH 7.4).

### 3.4. Predicting the injured muscle force

In the FIM the activation coefficient allows for the application of the impact of $Ca^{2+}$ ions, which plays a crucial role in disorders including the loss of force resulting from various muscular injuries.

Calcium ions play an important role in regulating the process of contraction and energy production. After binding $Ca^{2+}$ to troponin, they cause a conformational change, and the tropomyosin is pushed off, then actin sites become accessible to the myosin head. In the case of atrophy due to interruption of muscle nerve supply, calcium disturbance plays an important role in reducing force [47, 48].. The sarcoplasmic reticulum affinity to $Ca^{2+}$ in the denervated muscle is significantly lower than the intact muscle. The level of $Ca^{2+}$ ATPase activity in the SR, as well as ATP-dependent $Ca^{2+}$ binding, is also reduced [47]. Therefore, impaired intracellular calcium homeostasis is one of the important mechanisms causing atrophic disorder [48]. Hence, a lacerated muscle that is re-sutured to the healthy part may not be able to produce the same force as a control muscle due to atrophy [40]

As the ratio of free calcium in the sarcoplasm ($c$) to total calcium ($C$) decreases, the total activation coefficient decreases and muscle force decreases sharply (see Figures 8 and 9). This shows the capability of our FIM model to predict the effect of injury on muscle output.

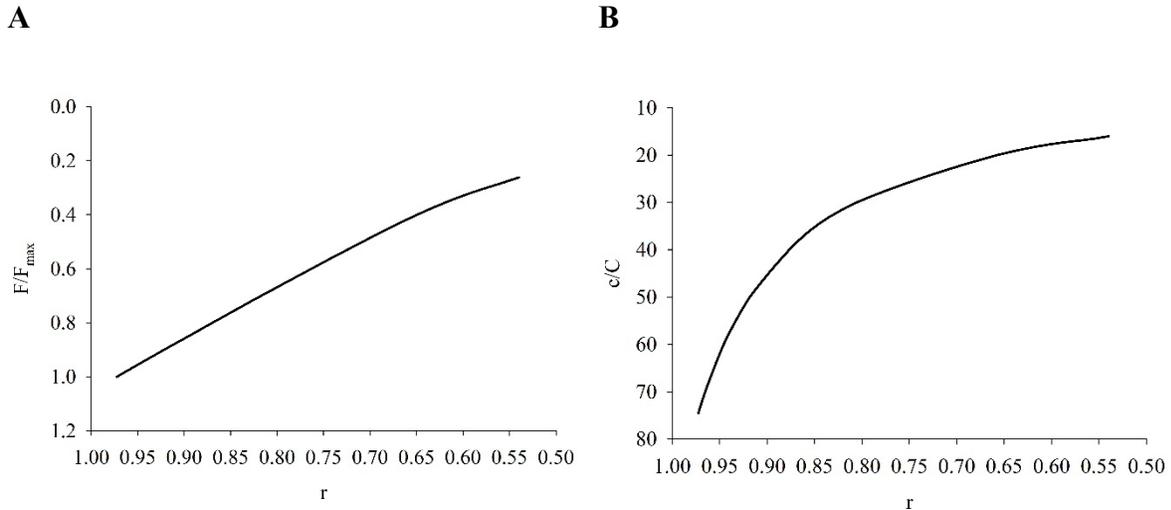

**Fig. 8. A:** The trend of normalized force changes in muscle with changes in activation coefficient, $r$. The activation coefficient depends both on the concentration of free calcium ions ($c$), which can bind to specific sites on troponin and make muscle contraction possible, and on the calcium-troponin equilibrium constant ($\mu$) (see Appendix). The activation coefficient has been used in the model to investigate the effect of participating calcium ($c$) in the contraction process. By reducing the activation coefficient from 98% to 52%, the muscle force is reduced about 80 percent. **B:** A graph of changes in the ratio of free calcium in the sarcoplasm ($c$) to total calcium ($C$), the sum of free calcium and calcium bound to troponin, versus the activation coefficient. The lower the proportion of free calcium to total calcium, the smaller the activation coefficient will be.

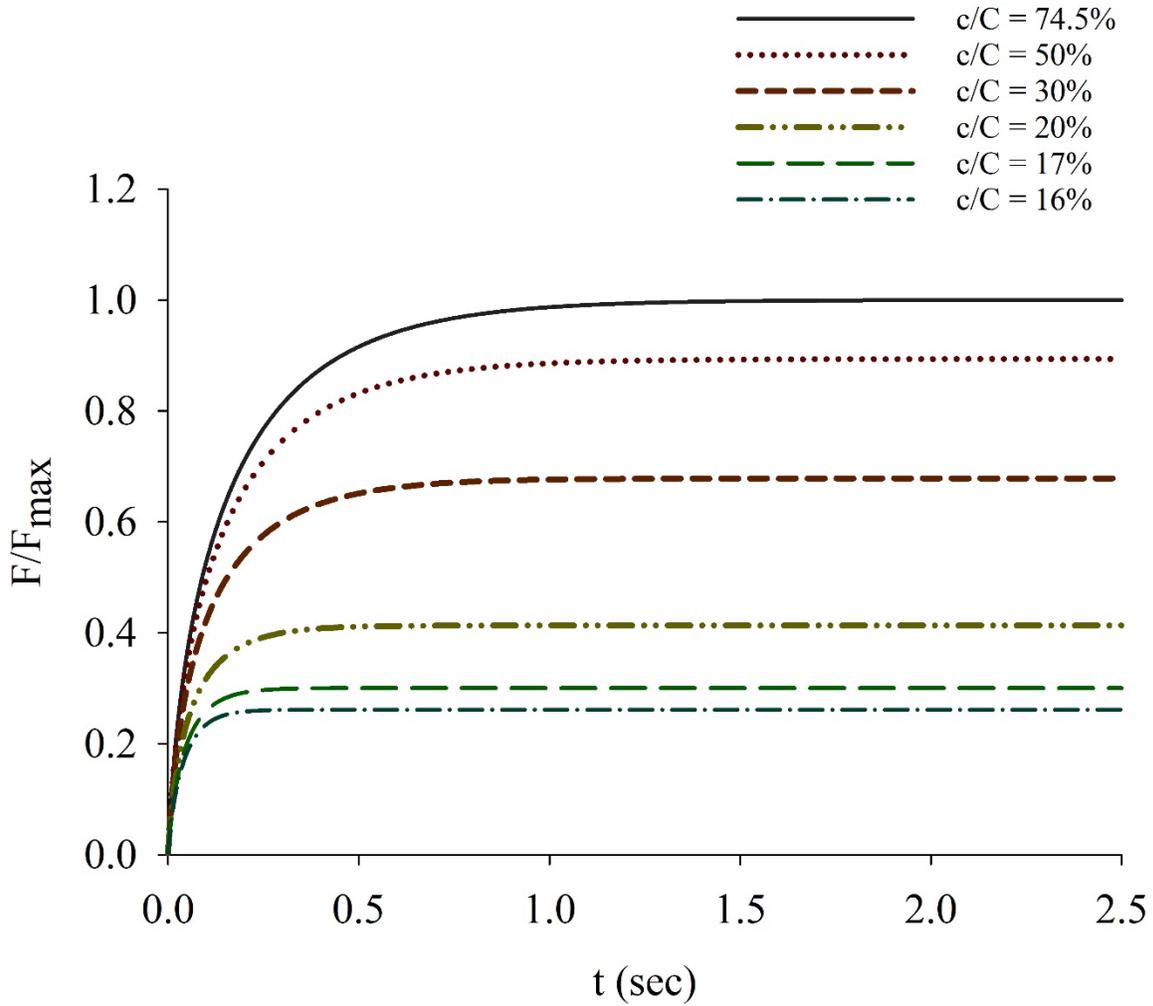

**Fig. 9.** Muscle force-time diagram. $c$ is free calcium in the sarcoplasm, which can react with troponin, and $C$ is total calcium in the myofibrillar space. The FIM model incorporates the activation coefficient ($r$), allowing the impact of calcium disruption to be incorporated into predicted muscle force. The ratio of free cytosolic calcium to total calcium is directly correlated to this activation coefficient. The sharp decline in the muscle force corresponds to a decrease in the total activation coefficient. In all diagrams, the tendon-muscle velocity is zero, and the muscle produces force due to stimulation from its initial frequency. The myosin transition rate from detached to strongly bound state ($f_a$), the ADP release rate ($f_D^0$), and the myosin rebinding rate to actin ($f_R$) are, 40 $s^{-1}$, 350 $s^{-1}$, and 0 $s^{-1}$, respectively. The detachment rates are the ones that have been reported in [34].

## 4. Discussion

In order to model the force of the fatigued muscle it was necessary to consider in detail the involved factors in muscle contraction. Studies have shown that among the contributing factors in muscle fatigue, the three ions of $Ca^{2+}$, $H^+$, and inorganic phosphate $P_i$ play a major role [21]. For this purpose, the following modifications were made to one of the most recent models, implemented according to the Distributed Moment approximation [34].

1. The effect of disturbances in the troponin-bound calcium, which is responsible for altering the structure of actin by moving tropomyosin filaments away from the actin sites, is considered in the model by defining a variable activation coefficient. With a specified ratio of free calcium to total calcium, the free calcium concentration is computed. Then, in contrast to a constant coefficient in the Zahalak model, the activation coefficient as a function of free calcium concentration is obtained.
2. Metabolic pathways of energy production are included in the model to consider the effect of $H^+$ and $P_i$ ions. It has been shown that the rate of myosin-to-actin binding from detached to strongly-bound state, the rate of ADP release from strongly bound myosin, and finally the myosin-rebinding rate to rigor state, all three depend on the concentration of $H^+$ in the muscle [17]. The first two rates decrease and the last rate increases with decreasing pH [17]. The reason behind these changes is that lowering the pH may reduce the number of force-generating crossbridges and also the force produced by each of them [29]. The binding rate of crossbridges in the DM energy model was replaced with the summation of the three above-mentioned rates and the rate of binding and separation of $P_i$ to the crossbridge in the rigor state.
3. Unlike acidosis, which completely affects the reactions related to the binding of actin to myosin, the main effect of increasing the $P_i$ is on the rate of myosin separation from actin. Hence the detachment rates were modified to include this effect.

Therefore, by modifying the attachment and the detachment rates of myosin to actin and also considering the activation factor, a more realistic model for muscle force was developed. The resulting FIM model can predict both the force of a healthy muscle and an anaerobic fatigued muscle.

The comparison between the maximum isometric force predicted by our FIM model and the one provided by the DM model shows that the force generated by the DM model is faster at the beginning and then both models converge to their corresponding isometric force at the optimum length. The replaced $f$ function is responsible for this change, which in turn is due to exponential attachment rates.

In the following, to investigate the FIM model outcome in fatigue conditions, first, the effect of increasing $H^+$ and then the simultaneous increase of $H^+$ and $P_i$ (30 mM) on the maximum isometric force of the muscle were studied. The percentage reduction in normalized force for a fatigued muscle is expected to be comparable to the experimental results. It was observed that the force is reduced by 9.5% at a physiological temperature due to the decrease in pH (Table 2). This value is in perfect agreement with the percentage reduction (4-12%) in isometric muscle force of the soleus and gastrocnemius rats reported by Knuth et al. [26], as well as the percentage reduction (about 10%) in the flexor brevis muscle force of the male mice [25]. Furthermore, the reported decrease in force level in [25] (10% force reduction), which has the most similar initial pH and temperature conditions as the FIM model, has a very good agreement (5% error) with the FIM model. However, it is less than the reported percentage reduction for rabbit's psoas muscle at pH 6.2 (17.8% reduction in force) [24]. The reason for this difference can be due to the type of fiber studied in rabbit muscle, which is in the category of fast fibers, and the pH value in the experiment which is lower than the pH related to the coefficients embedded in the model (Table 3). Then, by calculating the slope of the time-force diagram, the force generation and time constants were investigated. The force generation reaches 50.8% of its initial value due to the decrease in pH and the time constant increases by 72.7%. It can be seen that the error varies from 5.0% to 18.7%. This error can be explained by the fact that the amount of the initial and final pH in most experimental studies is different from the assumptions in the FIM model. Also, the attachment and detachment rates in the FIM belong to the fast twitch fiber types while the experiment in Knuth et al. [26] with higher error difference was done on a muscle containing fast and slow twitch fibers.

In the next step, to evaluate the model, the effect of 30 mM $P_i$ and the reduction of pH to 6.5 were simultaneously investigated. As shown, the force is reduced by 47.5% (Table 2). This reduction in force is quite consistent with the experimental results for slow and fast muscle fibers in rats (44% for slow 41, 46, and 50% for fast fibers) and is less than the 52% reduction in maximum isometric force reported for rabbit muscle, which is probably due to the type of fiber and the smaller amount of the pH, 6.2 instead of 6.5, in the study of rabbit muscle (Table 3). It can be seen that the error varies from 3.3% to 15.8%. This error can be explained by the fact that the amount of the initial and final pH in all experimental studies is different from the FIM model.

**Table 2.** Predicted results with the FIM model

|  | Fiber type | Temperature (°C) | Control Muscle pH | Fatigued Muscle pH | $P_i$ Concentration (mM) | Force Reduction (%) |
|---|---|---|---|---|---|---|
| FIM Prediction | Slow and Fast Twitch | 30 | 7.4 | 6.5 | 0 | 9.5 |
| FIM Prediction | Slow and Fast Twitch | 30 | 7.4 | 6.5 | 30 | 47.5 |

**Table 3.** Comparison of experimental results and predicted results with the FIM model

| | Fiber type | Temperature (°C) | Control Muscle pH | Fatigued Muscle pH | $P_i$ Concentration (mM) | Force Reduction (%) | Absolute Value of the Relative Error |
|---|---|---|---|---|---|---|---|
| **Knuth et al. 2006 [26]** | Fast and slow twitch | 30 | 7 | 6.2 | 0 | 4-12 | 0.187 |
| **Westerblad et al. 1997 [25]** | Fast twitch | 32 | 7.4 | 6.4 | 0 | 10 | 0.050 |
| **Karatzaferi et al. 2008 [30]** | Fast twitch | 30 | 7 | 6.2 | 30 | 52 | 0.086 |
| **Nelson et al. 2014 [31]** | Fast twitch | 30 | 7 | 6.2 | 30 | 46 | 0.033 |
| **Nelson et al. 2014 [32]** | Fast twitch | 30 | 7 | 6.2 | 30 | 50 | 0.050 |
| **Nelson et al. 2014 [32]** | Fast twitch | 30 | 7 | 6.2 | 30 | 41 | 0.158 |
| **Nelson et al. 2014 [32]** | Slow twitch | 30 | 7 | 6.2 | 30 | 44 | 0.079 |

In the developed model, it is possible to study the changes in the maximum isometric force in both the control muscle and the muscle in an acidosis state, in different shortening magnitudes and also for voluntary movements. As the shortening magnitudes increases, the overlap between myosin and actin filaments increases; this happens to the point that the muscle cannot produce force. As the shortening magnitude increases, the force decreases in both the control and fatigue states; the fatigue muscle force is less than the control muscle force in all the shortening magnitudes (Table 4). It is noteworthy to state that as the shortening magnitude increases, the initial slope does not change much, but the time constant increases. In voluntary movement, as expected, the muscle force decreases with the increasing speed of the tendon-muscle unit. The simultaneous increase in velocity and decrease in pH appear to cause fatigued muscle maximum isometric force to be reduced by more than 9.5%, which was the reduction percentage in muscle force at zero velocity. As the shortening velocity increases, the force decreases in both the control and fatigue states; the fatigue muscle force is less than the control muscle force at all the shortening velocities (Table 5).

**Table 4.** Normalized muscle force and its changes in control conditions and with a pH drop in different shortening magnitude

| Shortening Magnitude | $F/F_{max}$ | $\Delta F/F_{max}$ | $F/F_{max}$ | $\Delta F/F_{max}$ |
|---|---|---|---|---|
| | Control Muscle | | Fatigued Muscle | |
| 36 | 0.194 | 0.807 | 0.176 | 0.825 |
| 30 | 0.442 | 0.559 | 0.399 | 0.601 |
| 27 | 0.548 | 0.452 | 0.494 | 0.506 |
| 24 | 0.643 | 0.357 | 0.579 | 0.421 |
| 15 | 0.861 | 0.139 | 0.772 | 0.229 |
| 0 | 1 | 0 | 0.905 | 0.095 |

**Table 5.** Normalized muscle force and its changes in control conditions and with a pH drop at different shortening velocities

| Shortening velocity | $F/F_{max}$ | $\Delta F/F_{max}$ | $F/F_{max}$ | $\Delta F/F_{max}$ |
|---|---|---|---|---|
| | Control Muscle | | Fatigued Muscle | |
| 0 | 1 | 0 | 0.906 | 0.094 |
| 0.5 | 0.827 | 0.1737 | 0.568 | 0.432 |
| 1.5 | 0.621 | 0.3786 | 0.335 | 0.665 |

Calcium has a key role in skeletal muscle contraction, and it is expected that disturbances in the concentration of $Ca^{2+}$ also play a significant role in reducing muscle force. Although $H^+$ and $P_i$ play an important role in muscle fatigue, impaired $Ca^{2+}$ content and function have been seen in addition to fatigue in a wider range of muscle disorders, such as atrophy, dystrophy, and denervation. The effect of calcium ions can be investigated by using the activation factor in such a way that as the activation factor decreases, the number of cross-bridges that interact with the actin site and produce energy decreases. This is considering the catabolic effect of $Ca^{2+}$ on muscle contractile proteins. In this coefficient (activation factor), the reaction rate of $Ca^{2+}$ binding to troponin, the reaction rate of $Ca^{2+}$ separation from troponin, and the concentration of free $Ca^{2+}$ in the sarcoplasm are related. It was assumed that all free calcium in the sarcoplasm could react with troponin to evaluate the effect of the activation coefficient on muscle force. Since 65 to 84% of the calcium in the sarcoplasm has been shown to interact with troponin [46], for the initial ratio of free calcium to total calcium in the myofibrillar space the average of these two values was assumed i.e. 74.5%. As expected, muscle force was significantly reduced by reducing the calcium available for troponin binding reaction, when only 15% of the total calcium in the myofibrillar space was able to interact with troponin, muscle force was reduced up to 80% (Fig. 8, Fig. 9). In addition, by examining the time force graphs for different ratios of free calcium in the sarcoplasm to total calcium, it is observed that the slope of the graphs does not change (Fig. 9).

One of the advantages of DM-based models is their ability to calculate the stored elastic energy in cross-bridges. Power is energy per unit of time, and by taking the second moment of the moment distribution function, an understanding of muscle power can be achieved. Since tendon stiffness is used in moment equations, the effect of muscle-tendon stiffness also is taken into account in the calculation of elastic energy. In a work studying the effect of reducing the pH on type II fibers of rats, it was observed that the maximum power is reduced by 18 to 34% [26]. Having computed the maximum energy produced by the muscle, the maximum muscle power decreases by 18% due to a decrease in pH from 7.4 to 6.5 which shows that the FIM model's prediction is consistent with experimental data (Fig. 10). Due to the simultaneous decrease of pH and increase of $P_i$ concentration, the muscle can produce approximately 50% of its previous power, which is in agreement with the reported inhibited power (55%) in the rabbit's psoas muscle (with a relative error value of 10%) [30].

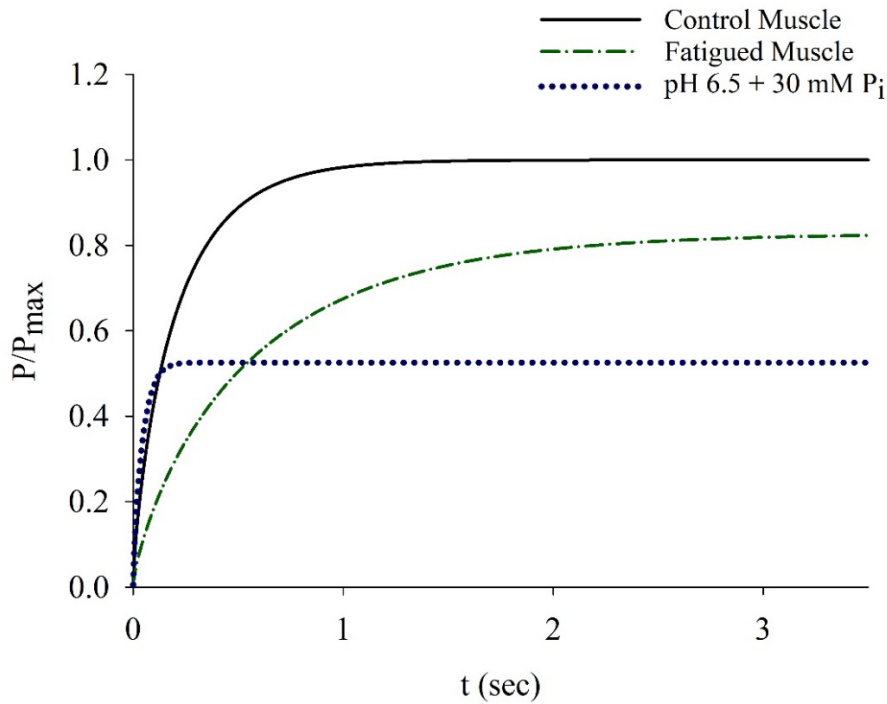

**Fig. 10.** Predicting changes in muscle power in three modes: control muscle (solid line), muscle in fatigue conditions due to accumulation of hydrogen ions (dash-dotted line), and muscle in which the pH decreases and the $P_i$ concentration increases to 30 mM simultaneously (dotted line). The values of the myosin transition rate from detached to strongly bound state ($f_a$), ADP release rate ($f_D^0$), and myosin rebinding rate to actin ($f_R$) for a healthy muscle are $40\ s^{-1}$, $350\ s^{-1}$, and $0\ s^{-1}$, respectively [17]. The detachment rate values are according to [34]. For the fatigued muscle due to a drop in pH, the corresponding rates of $10\ s^{-1}$, $100\ s^{-1}$, and $10\ s^{-1}$ for myosin transition from detached to strongly bound state, ADP release, and myosin rebinding to actin, respectively, are used [17]. The values of the detachment rate are similar to those for a healthy muscle (control muscle) [34]. Due to a pH drop the normalized power reduces by 18%. The rates for a muscle subjected to a pH drop from 7.4 to 6.5 and simultaneous inorganic phosphate accumulation (30 mM) are as follows [17]: $f_a = 10\ s^{-1}$, $f_D^0 = 100\ s^{-1}$, $f_R = 10\ s^{-1}$, $g_{off} = 100\ mM^{-1}\ s^{-1}$, $f_P^+ = 0.7\ mM^{-1}\ s^{-1}$, $f_P^- = 10\ s^{-1}$, and $g_{DP} = 40\ s^{-1}$. The rate at which myosin detaches from actin due to the binding of $[P_i]$ when myosin is strongly bound to actin is referred to as $g_{off}$. The rate at which $P_i$ and ADP are released from the active site of myosin when it is not bound to actin is known as $g_{DP}$. Furthermore, $f_P^+$ denotes the reaction rate for the release of inorganic phosphate from myosin while it is bound to actin, and $f_P^-$ represents the reverse rate of this reaction (see text for more details). The simultaneous accumulation of hydrogen and inorganic phosphate ions causes a 50% reduction in normalized power.

Muscle stiffness is an indicator for estimating attached crossbridges. To evaluate the stiffness of muscle fibers, it is a custom to compute force changes in response to sinusoidal length oscillations at different frequencies [49-55]. In [49], by applying tetanus tension to muscle over short periods, muscle fatigue in the tibial frog muscle was examined. The authors found that in moderate fatigue, the force and stiffness of muscle both decrease, but the reduction in force is far larger than the reduction in stiffness. In another study [50] on both fast- and slow-twitch muscle fibers, similar results were found for rabbit vastus lateralis muscle fiber. It was found that the stiffness in slow-twitch muscle fiber did not change with a decrease in pH at both 10 and 15 °C temperatures. The results obtained from the effect of increasing the concentration of inorganic phosphate on muscle stiffness are different. [51-53, 55, 56] have reported a decrease in force and muscle stiffness with an increasing concentration of inorganic phosphate. The decrease in stiffness resulting from increasing the concentration of this ion has been reported less than the decrease in force. For example, in the study of the rabbit psoas muscle [52], it was observed that by increasing the concentration of $P_i$ by photolysis of 1- (2-nitrophenyl) ethyl phosphate, the isometric force of the muscle decreases. In addition to the decrease in isometric force, muscle stiffness also decreases. The stiffness change has been reported as about 83% of the force change. The stiffness drop has been attributed to decreasing number of high-force crossbridges due to the binding of inorganic phosphate ions to the crossbridge.

As mentioned at the beginning of the study, Huxley-based models, in addition to the ability to model the force and elastic energy of the entire muscle, can display muscle stiffness. In the FIM model, the stiffness related to the optimum length is studied in the plateau section of the contraction curve. Change in muscle stiffness due to increased concentration of hydrogen ions

(6.5% stiffness reduction) is consistent with the results of previous studies [49, 50]. The result of the simultaneous increase in the concentration of $H^+$ and $P_i$ ions (40% stiffness reduction) is consistent with [51-53, 56].

In comparison to previous models, our model takes into account a multitude of factors that were not considered simultaneously in one model previously. Unlike microscopic branched pathway [17] and Walcott [36] models, which focused solely on actin-myosin interaction to simulate muscle contraction, we have incorporated additional variables such as tendon compliance ($K(Q_1)$), the degree of overlap between actin and myosin ($\alpha$), the influence of regulatory proteins like troponin, and the role of $Ca^{2+}$; which is one of the most important ions not only in muscle contraction but also in its excitation.

The FIM model is also a significant improvement over many of the previous Huxley-based models [7, 8, 10, 34] due to its consideration of the essential role of calcium ions in the excitation-contraction process. This model is capable of accurately predicting muscle stiffness, force, and energy in various conditions, including healthy and fatigued muscles affected by $H^+$ and $P_i$. One of the unique features of the FIM model is its ability to demonstrate the negative impact of calcium ions, $H^+$, and $P_i$ on muscle contraction, which no other Huxley-based model [7-11, 34] has been able to achieve.

While our FIM model is admittedly more complex than Hill-type models, the ones that consider the role of calcium [3, 4, 57-59] and those that do not [60, 61], it remains more streamlined than previous microscopic fatigue models since it consists of only four coupled, first-ordinary differential equations that can easily be solved using the distribution moment method.

## 5. Limitations

After a severe laceration injury in skeletal muscles, one of the certain events is the activation of fibro-adipogenic progenitors cells. These cells differentiate into fibroblasts and adipocytes [62]; thus, they cause force reduction since the newly differentiated cells are not like active muscle cells. In phenomenological muscle models like the Hill model, the parallel elastic element represents fascia and connective tissue, but there is not such an element in the Huxley model; thus, the effect of these passive tissues in muscle force was not studied in the FIM model. Also, it is necessary to note that the connective tissues' resistance is larger in eccentric contractions than concentric ones; therefore, we predict that this fibrogenesis does not alter the concentric contractions results significantly.

In addition, sometimes after regeneration buds are formed from the old muscle fibers; subsequently, the regenerated cells are not the same as the original cells. The regenerated cells are generally smaller than the original cells [40]. These changes in muscle fibers can cause an alteration in the distance between two consecutive binding sites on actin and ultimately cause a change in the optimal length of the muscle. In this case, in addition to changing the rates and the activation coefficient, the changes in $h$, the maximum distance that a cross-bridge can move to make contact with the active site, and $l$, the distance between two consequent actin sites, should also be studied.

# 6. Conclusion

Muscle modeling has been strikingly growing since it has a vital role in understanding movement production. Although a number of muscle models have been proposed, many challenges still exist in presenting a model that captures how different impairments affect muscle force. Achieving such a model necessitates adding new variables to a healthy muscle model which can consider the effects of impairments Studying various muscle injuries, including fatigue due to short but intense activity, fatigue through long-term activities, atrophy, denervation, and partial as well as complete laceration, we concluded that despite that each muscle injury has a unique and complex mechanism, there are two common consequences: force and power loss. A damaged muscle is not able to make as much force and power as a healthy muscle. This led us to find the most effective variables that account for this phenomenon in producing force and power in skeletal muscles.

The Huxley equation was used as the base of the FIM model. To evaluate metabolic pathways, modifying actin-myosin attachment and detachment rates was required. Therefore, new myosin and actin-attachment rates were defined, showing the sum of the fundamental force- and displacement-dependent reaction rates in the muscle contraction cycle. Also, the contractile proteins' detachment rates were modified. Finally, an activation coefficient was defined in the model representing the ratio of calcium ions that interact with troponin and produce muscle force. These modifications made a framework to study the effect of three of the most crucial ions in force and power generation, i.e., $H^+$, $P_i$, and $Ca^{2+}$.

The FIM model predicted results for muscle force reduction in fatigue conditions due to pH reduction (9.5% force reduction) and as a consequence of simultaneous pH reduction and $P_i$ addition (47.5% force reduction), which are compatible with the experimental results of multiple previous studies. The FIM model also enables us to investigate the effect of the varying initial shortening magnitudes as well as shortening velocity. In the following the muscle force in an injured muscle due to $Ca^{2+}$ ions dysfunction was predicted. By changing the ratio of participating $Ca^{2+}$ ions to total $Ca^{2+}$ ions and calculating the activation coefficient the injured muscle force is predicted. The muscle force falls dramatically due to a decline in the activation coefficient which induces a reduction in the number of attached crossbridges. As a further application, the muscle stiffness and muscle power were obtained in two cases by the FIM. The first case was an increase in the concentration of hydrogen ions, and the second case was a simultaneous increase in the concentration of $H^+$ ions and $P_i$. The obtained results for changes in muscle stiffness and power in both mentioned states were in acceptable agreement with the experimental results. We can say that the developed model is a flexible model that can display muscle force, power, and stiffness in a wide range of states such as fatigue, atrophy, neuromuscular defects, and eventually injury.

## Appendix

The basis of the relations of the FIM model is similar to the modified Huxley model [10], but in order to provide more information, a complete explanation of the equations of the model is given.

The first step is to estimate the new $f$. For this purpose, polynomials with degrees 3 and 5 are used. The criterion for selecting the degree of polynomials is the lowest possible degree with high accuracy. As mentioned earlier the forward binding rate is considered as follows,

$$f(x, F, [P_i]) = f_a(x) + f_D(F) + f_R(x) + (f_P^+ - f_P^-) \quad \text{(A.1)}$$

$$f_a(x) = f_a \sqrt{\frac{kh^2}{2\pi K_B T}} \exp\left(-\frac{kx^2}{2K_B T}\right) \quad \text{(A.2) [12]}$$

where $k = 0.3\ pN/nm$ is myosin's stiffness, $K_B$ is Boltzmann's constant, T=303 Kelvin is absolute temperature, x is the displacement of cross-bridge from its equilibrium position, and $f_a$ is the initial value of $f_a(x)$. First, all bond rates must be normalized using the $h$ parameter. Assuming $\xi = x/h$, the normalized form of Equation A.2 is as follows,

$$f_a(\xi) = f_a \sqrt{\frac{kh^2}{2\pi K_B T}} \exp\left(-\frac{k\xi^2 h^2}{2K_B T}\right) \quad \text{(A.3)}$$

Using the curve fitting tool in MATLAB software, it was found that a fifth-degree polynomial can represent this exponential relationship with acceptable accuracy ($R - square = 0.9944 \cong 1$). Thus the $f_a(\xi)$ equation is replaced by the following polynomial in which the $A_j s$ are fixed polynomial coefficients.

$$f_a(\xi) = f_a(A_5\xi^5 + A_4\xi^4 + A_3\xi^3 + A_2\xi^2 + A_1\xi + A_0) \tag{A.4}$$

With a similar process, the $f_R(\xi)$ equation is as follows, in which $f_R$ is the initial value of rebinding rate to rigor state,

$$f_R(\xi) = f_R(A_5\xi^5 + A_4\xi^4 + A_3\xi^3 + A_2\xi^2 + A_1\xi + A_0) \tag{A.5}$$

The $f_D(F)$ equation is as follows:

$$f_D(F) = f_D^0 \exp\left(-\frac{F\varepsilon_x}{K_B T}\right) \tag{A.6}$$

where $f_D^0$ is the ATP release rate when the force is zero, $\varepsilon_x = 1.86\ nm$ is the distance from the transition state, assuming a linear equation for the force, (A.3) becomes the following relation, in which d=10 nm is the myosin's step size.

$$f_D(x) = f_D^0 \exp\left(-\frac{k(x+d)\varepsilon_x}{K_B T}\right)$$
$$f_D(\xi) = f_D^0 \exp\left(-\frac{kd\varepsilon_x}{K_B T}\right) \exp\left(-\frac{kh\varepsilon_x}{K_B T}\xi\right) \tag{A.7}$$

By fitting the curve to the exponential part of the $f_D(\xi)$ function (A.7), it can be seen that a third order polynomial fits perfectly on the curve ($R - square = 0.9992 \cong 1$). Thus, the $f_D(\xi)$ equation is replaced by the following polynomial in which the $D_j s$ are fixed polynomial coefficients.

$$f_D(\xi) = f_D^0(D_3\xi^3 + D_2\xi^2 + D_1\xi + D_0) \tag{A.8}$$

Tendon compliance, which is estimated as a variable value of the first moment of the bond distribution, is as follows ((A.9)). In (A.9), $E_{max}$ is the amount of tendon strain when exposed to the maximum isometric force means, $F_{max}$, $L_{slack}$ is the slack length of the tendon, $L_{M\_T(opt)}$ is the optimum length of the muscle-tendon complex, $\Gamma$ is a constant value that relates the first moment of the bond distribution function to force [10].

$$K(Q_1) = \frac{E_{max} L_{slack}}{2 L_{M\_T(opt)}} \sqrt{\frac{\Gamma}{F_{max} Q_1}} \qquad (A.9)$$

Depending on the degree of actin and myosin overlap, the number of cross-bridges that can bind to the actin site varies [34]. This coefficient of overlap is displayed with $\alpha$ which depends on the current length of the muscle-tendon unit, $L_{M\_T}$. The $\alpha$ relation is extracted from the force-length experimental diagram [35].

$$\alpha = \alpha_1 \left(\frac{L_{M\_T}}{L_{M\_T(opt)}}\right)^2 + \alpha_2 \frac{L_{M\_T}}{L_{M\_T(opt)}} + \alpha_3 \qquad (A.10)$$

where $\alpha_1$, $\alpha_2$, and $\alpha_3$ are -6.25, 12.5. and -5.25 respectively.

The parameters used in the model are derived from the three articles [11], [34], and [17]. Some of these parameters were listed in Table 1 and the rest is given in Table 1.A.

**Table 1.A.** Parameter values used in the model

| $E_{max}$ | $F_{max}$ | $L_{slack}$ | $L_{M\_T(opt)}$ | $\Gamma$ | $L_{C\_E(opt)}$ | h | $s_0$ | $\varepsilon_0$ | $\mu$ | b |
|---|---|---|---|---|---|---|---|---|---|---|
| 0.024 | 1 | 41 mm | 100 mm | 1.7898 | 32.57 | 27 nm | 2.4 μm | 1.11 | 0.1 | 0.5 |

$E_{max}$ is the amount of tendon nominal strain when exposed to the maximum isometric force, i.e., $F_{max}$. $L_{slack}$ is the slack length of the tendon, $L_{M\_T(opt)}$ is the optimum length of the muscle-tendon, $L_{C\_E(opt)}$ is the contractile element length when the muscle-tendon length is optimum, $\Gamma$ is a constant value that relates the first moment of the bond distribution function to force, $s_0$ is sarcomere length in reference condition, $\mu$ is Calcium-troponin equilibrium constant, b is cross-bridge structural parameter, and $\varepsilon_0$ is myosin bond length at yield normalized by h. Parameters without units are dimensionless.

By considering the activation coefficient and $\alpha$, the DM model is transformed into a system of four coupled ordinary differential equations.

$$u(t) = \frac{L_{M_T} s_0}{L_{M_{T(opt)}} h} \left(\dot{Q}_1 K(Q_1) - \frac{\dot{L}_{M\_T}}{L_{M\_T(opt)}}\right) \qquad (A.11)$$

$$\dot{Q}_0 = r(c)\alpha\beta_0 - r(c)\phi_{10} - \phi_{20} \qquad (A.12)$$

$$\dot{Q}_1 = r(c)\alpha\beta_1 - r(c)\phi_{11} - \phi_{21} - u(t)Q_0 \qquad (A.13)$$

$$\dot{Q}_2 = r(c)\alpha\beta_2 - r(c)\phi_{12} - \phi_{22} - 2u(t)Q_1 \qquad (A.14)$$

with

$$C(c, Q_0) = c + 2bQ_0 + r(c)\left(2 + \frac{\mu}{c}\right)(1 - bQ_0) \qquad (A.15)$$

$$r(c) = \frac{c^2}{c^2 + \mu c + \mu^2} \qquad (A.16)$$

where *u(t)* is the myofilament's shortening velocity normalized by h, $\beta_i$s are the attachment rate's moments, $\phi_i$s are the function of both attachment and detachment rates, and bond distribution. $s_0$ is sarcomere length at reference condition, *c* is cytosol calcium concentration which bind to troponin and produce force, $\mu$ is calcium-troponin equilibrium constant, *b* is cross-bridge structural parameter, and *C* is total calcium concentration. The dot symbol above *Q* represents a derivative with respect to time.

$$\phi_{1i} = Q_0 \Bigg( (f_a + f_R) A_5 \int_0^1 \frac{1}{\sqrt{2\pi}q} \xi^{i+5} e^{\frac{-(\xi-p)^2}{2q^2}} d\xi$$

$$+ (f_a + f_R) A_4 \int_0^1 \frac{1}{\sqrt{2\pi}q} \xi^{i+4} e^{\frac{-(\xi-p)^2}{2q^2}} d\xi$$

$$+ (f_D^0 D_3 + (f_a + f_R) A_3) \int_0^1 \frac{1}{\sqrt{2\pi}q} \xi^{i+3} e^{\frac{-(\xi-p)^2}{2q^2}} d\xi + (f_D^0 D_2$$

$$+ (f_a + f_R) A_2) \int_0^1 \frac{1}{\sqrt{2\pi}q} \xi^{i+2} e^{\frac{-(\xi-p)^2}{2q^2}} d\xi \quad \text{(A.17)}$$

$$+ (f_D^0 D_1 + (f_a + f_R) A_1) \int_0^1 \frac{1}{\sqrt{2\pi}q} \xi^{i+1} e^{\frac{-(\xi-p)^2}{2q^2}} d\xi + (f_D^0 D_0$$

$$+ (f_a + f_R) A_0 + (f_P^+ - f_P^-)) \int_0^1 \frac{1}{\sqrt{2\pi}q} \xi^i e^{\frac{-(\xi-p)^2}{2q^2}} d\xi \Bigg)$$

$$\phi_{2i} = Q_0 (g_2 \int_{-\infty}^0 \frac{1}{\sqrt{2\pi}q} \xi^i e^{\frac{-(\xi-p)^2}{2q^2}} d\xi$$

$$+ g_1 \int_0^1 \frac{1}{\sqrt{2\pi}q} \xi^{i+1} e^{\frac{-(\xi-p)^2}{2q^2}} d\xi + (g_1$$

$$+ g_3) \int_1^\infty \frac{1}{\sqrt{2\pi}q} \xi^{i+1} e^{\frac{-(\xi-p)^2}{2q^2}} d\xi \quad \text{(A.18)}$$

$$- g_3 \int_1^\infty \frac{1}{\sqrt{2\pi}q} \xi^i e^{\frac{-(\xi-p)^2}{2q^2}} d\xi)$$

where $p = \frac{Q_1}{Q_0}$ and $q = \sqrt{\frac{Q_2}{Q_0} - (\frac{Q_1}{Q_0})^2}$ (A19).

If the integral J is as follows,

$$J_l(\eta) = \frac{1}{\sqrt{2\pi}q} \int_{-\infty}^\eta \xi^l e^{\frac{-(\xi-p)^2}{2q^2}} d\xi \quad \text{(A.20)}$$

using two variable conversions $\zeta = \frac{\xi-p}{q}$ and $\tau = \frac{\eta-p}{q}$, $J_l(\eta)$ becomes the following integral.

$$J_l(\tau) = \frac{1}{\sqrt{2\pi}} \int_{-\infty}^{\tau} (p + q\zeta)^l e^{\frac{-\zeta^2}{2}} d\zeta \qquad (A.21)$$

by the definition of $\psi(\tau)$, as follows

$$\psi(\tau) = \frac{1}{\sqrt{2\pi}} \int_{-\infty}^{\tau} e^{\frac{-\zeta^2}{2}} d\zeta = \frac{1}{2} \text{erf}\left(\frac{\tau}{\sqrt{2}}\right) + \frac{1}{2} \qquad (A.22)$$

finally, the $J_l(\tau)$ functions will be in the form of a set of equations (A.23).

$$J_0(\tau) = \psi(\tau)$$

$$J_1(\tau) = p\psi(\tau) - q \frac{e^{\frac{-\tau^2}{2}}}{\sqrt{2\pi}}$$

$$J_2(\tau) = p^2\psi(\tau) - 2pq \frac{e^{\frac{-\tau^2}{2}}}{\sqrt{2\pi}} + q^2 \left(\psi(\tau) - \tau \frac{e^{\frac{-\tau^2}{2}}}{\sqrt{2\pi}}\right)$$

$$J_3(\tau) = p^3\psi(\tau) - 3p^2q \frac{e^{\frac{-\tau^2}{2}}}{\sqrt{2\pi}} + 3pq^2 \left(\psi(\tau) - \tau \frac{e^{\frac{-\tau^2}{2}}}{\sqrt{2\pi}}\right)$$

$$- q^3 \left((2 + \tau^2) \frac{e^{\frac{-\tau^2}{2}}}{\sqrt{2\pi}}\right) \qquad (A.23)$$

$$J_4(\tau) = p^4\psi(\tau) - 4p^3q \frac{e^{\frac{-\tau^2}{2}}}{\sqrt{2\pi}} + 6p^2q^2 \left(\psi(\tau) - \tau \frac{e^{\frac{-\tau^2}{2}}}{\sqrt{2\pi}}\right)$$

$$- 4pq^3 \left((2 + \tau^2) \frac{e^{\frac{-\tau^2}{2}}}{\sqrt{2\pi}}\right) + q^4 \left(3\psi(\tau) - \tau(\tau^2 + 3) \frac{e^{\frac{-\tau^2}{2}}}{\sqrt{2\pi}}\right)$$

$$J_5(\tau) = p^5\psi(\tau) - 5p^4q\frac{e^{\frac{-\tau^2}{2}}}{\sqrt{2\pi}} + 10p^3q^2\left(\psi(\tau) - \tau\frac{e^{\frac{-\tau^2}{2}}}{\sqrt{2\pi}}\right)$$

$$- 10p^2q^3\left((2+\tau^2)\frac{e^{\frac{-\tau^2}{2}}}{\sqrt{2\pi}}\right)$$

$$+ 5pq^4\left(3\psi(\tau) - \tau(\tau^2+3)\frac{e^{\frac{-\tau^2}{2}}}{\sqrt{2\pi}}\right)$$

$$- q^5\left((\tau^4 + 4\tau^2 + 8)\frac{e^{\frac{-\tau^2}{2}}}{\sqrt{2\pi}}\right)$$

$$J_6(\tau) = p^6\psi(\tau) - 6p^5q\frac{e^{\frac{-\tau^2}{2}}}{\sqrt{2\pi}} + 15p^4q^2\left(\psi(\tau) - \tau\frac{e^{\frac{-\tau^2}{2}}}{\sqrt{2\pi}}\right)$$

$$- 20p^3q^3\left((2+\tau^2)\frac{e^{\frac{-\tau^2}{2}}}{\sqrt{2\pi}}\right)$$

$$+ 15p^2q^4\left(3\psi(\tau) - \tau(\tau^2+3)\frac{e^{\frac{-\tau^2}{2}}}{\sqrt{2\pi}}\right)$$

$$- 6pq^5\left((\tau^4 + 4\tau^2 + 8)\frac{e^{\frac{-\tau^2}{2}}}{\sqrt{2\pi}}\right)$$

$$+ q^6\left(15\psi(\tau) - \tau(\tau^4 + 5\tau^2 + 15)\frac{e^{\frac{-\tau^2}{2}}}{\sqrt{2\pi}}\right)$$

$$J_7(\tau) = p^7\psi(\tau) - 7p^6q\frac{e^{\frac{-\tau^2}{2}}}{\sqrt{2\pi}} + 21p^5q^2\left(\psi(\tau) - \tau\frac{e^{\frac{-\tau^2}{2}}}{\sqrt{2\pi}}\right)$$

$$- 35p^4q^3\left((2+\tau^2)\frac{e^{\frac{-\tau^2}{2}}}{\sqrt{2\pi}}\right)$$

$$+ 35p^3q^4\left(3\psi(\tau) - \tau(\tau^2+3)\frac{e^{\frac{-\tau^2}{2}}}{\sqrt{2\pi}}\right)$$

$$- 21p^2q^5\left((\tau^4+4\tau^2+8)\frac{e^{\frac{-\tau^2}{2}}}{\sqrt{2\pi}}\right)$$

$$+ 7pq^6\left(15\psi(\tau) - \tau(\tau^4+5\tau^2+15)\frac{e^{\frac{-\tau^2}{2}}}{\sqrt{2\pi}}\right)$$

$$- q^7\left((\tau^6+6\tau^4+24\tau^2+48)\frac{e^{\frac{-\tau^2}{2}}}{\sqrt{2\pi}}\right)$$

By replacing the set of equations (A.23) in equations (A.17) and (A.18), functions $\phi_{1i}$, and $\phi_{2i}$ are determined ((A.24)).

$$\phi_{20} = Q_0\left(g_2 J_0\left(\frac{-p}{q}\right) + (g_1 + 0.5 g_{off})\left(J_1\left(\frac{1-p}{q}\right) - J_1\left(\frac{-p}{q}\right)\right)\right.$$

$$+ \left(g_1 + g_3 + (c_{P_i} - 0.5) g_{off} + g_{DP}\right)\left(p - J_1\left(\frac{1-p}{q}\right)\right) \quad (A.24)$$

$$\left. - \left(g_3 + (c_{P_i} - 0.5) g_{off} + g_{DP}\right)\left(1 - J_0\left(\frac{1-p}{q}\right)\right)\right)$$

$$\phi_{21} = Q_0 \left( g_2 J_1 \left( \frac{-p}{q} \right) + (g_1 + 0.5 g_{off}) \left( J_2 \left( \frac{1-p}{q} \right) - J_2 \left( \frac{-p}{q} \right) \right) \right.$$

$$+ (g_1 + g_3 + (c_{P_i} - 0.5) g_{off} + g_{DP}) \left( p^2 + q^2 \right.$$

$$\left. - J_2 \left( \frac{1-p}{q} \right) \right) - (g_3$$

$$\left. + (c_{P_i} - 0.5) g_{off} + g_{DP}) \left( p - J_1 \left( \frac{1-p}{q} \right) \right) \right)$$

$$\phi_{22} = Q_0 \left( g_2 J_2 \left( \frac{-p}{q} \right) + (g_1 + 0.5 g_{off}) \left( J_3 \left( \frac{1-p}{q} \right) - J_3 \left( \frac{-p}{q} \right) \right) \right.$$

$$+ (g_1 + g_3 + (c_{P_i} - 0.5) g_{off} + g_{DP}) \left( p^3 + 3pq^2 \right.$$

$$\left. - J_3 \left( \frac{1-p}{q} \right) \right)$$

$$\left. - (g_3 + (c_{P_i} - 0.5) g_{off} + g_{DP}) \left( p^2 + q^2 - J_2 \left( \frac{1-p}{q} \right) \right) \right)$$

$$\phi_{10} = Q_0 \left( ((f_a + f_R) A_0 + f_D^0 D_0 + (f_P^+ - f_P^-)) \left( J_0 \left( \frac{1-p}{q} \right) - J_0 \left( \frac{-p}{q} \right) \right) \right.$$

$$+ ((f_a + f_R) A_1 + f_D^0 D_1) \left( J_1 \left( \frac{1-p}{q} \right) - J_1 \left( \frac{-p}{q} \right) \right)$$

$$+ ((f_a + f_R) A_2 + f_D^0 D_2) \left( J_2 \left( \frac{1-p}{q} \right) - J_2 \left( \frac{-p}{q} \right) \right)$$

$$+ ((f_a + f_R) A_3 + f_D^0 D_3) \left( J_3 \left( \frac{1-p}{q} \right) - J_3 \left( \frac{-p}{q} \right) \right)$$

$$+ ((f_a + f_R) A_4) \left( J_4 \left( \frac{1-p}{q} \right) - J_4 \left( \frac{-p}{q} \right) \right)$$

$$\left. + ((f_a + f_R) A_5) \left( J_5 \left( \frac{1-p}{q} \right) - J_5 \left( \frac{-p}{q} \right) \right) \right)$$

$$\phi_{11} = Q_0 \Bigg( \big((f_a + f_R)A_0 + f_D^0 D_0 + (f_P^+ - f_P^-)\big)\left(J_1\left(\frac{1-p}{q}\right) - J_1\left(\frac{-p}{q}\right)\right)$$

$$+ \big((f_a + f_R)A_1 + f_D^0 D_1\big)\left(J_2\left(\frac{1-p}{q}\right) - J_2\left(\frac{-p}{q}\right)\right)$$

$$+ \big((f_a + f_R)A_2 + f_D^0 D_2\big)\left(J_3\left(\frac{1-p}{q}\right) - J_3\left(\frac{-p}{q}\right)\right)$$

$$+ \big((f_a + f_R)A_3 + f_D^0 D_3\big)\left(J_4\left(\frac{1-p}{q}\right) - J_4\left(\frac{-p}{q}\right)\right)$$

$$+ \big((f_a + f_R)A_4\big)\left(J_5\left(\frac{1-p}{q}\right) - J_5\left(\frac{-p}{q}\right)\right)$$

$$+ \big((f_a + f_R)A_5\big)\left(J_6\left(\frac{1-p}{q}\right) - J_6\left(\frac{-p}{q}\right)\right) \Bigg)$$

$$\phi_{12} = Q_0 \Bigg( \big((f_a + f_R)A_0 + f_D^0 D_0 + (f_P^+ - f_P^-)\big)\left(J_2\left(\frac{1-p}{q}\right) - J_2\left(\frac{-p}{q}\right)\right)$$

$$+ \big((f_a + f_R)A_1 + f_D^0 D_1\big)\left(J_3\left(\frac{1-p}{q}\right) - J_3\left(\frac{-p}{q}\right)\right)$$

$$+ \big((f_a + f_R)A_2 + f_D^0 D_2\big)\left(J_4\left(\frac{1-p}{q}\right) - J_4\left(\frac{-p}{q}\right)\right)$$

$$+ \big((f_a + f_R)A_3 + f_D^0 D_3\big)\left(J_5\left(\frac{1-p}{q}\right) - J_5\left(\frac{-p}{q}\right)\right)$$

$$+ \big((f_a + f_R)A_4\big)\left(J_6\left(\frac{1-p}{q}\right) - J_6\left(\frac{-p}{q}\right)\right)$$

$$+ \big((f_a + f_R)A_5\big)\left(J_7\left(\frac{1-p}{q}\right) - J_7\left(\frac{-p}{q}\right)\right) \Bigg)$$